\documentstyle[epic,eepic,aps,prd]{revtex}

\begin{document}
\draft   

\title{Retarded Greens Functions and Forward Scattering Amplitudes in
Thermal Field Theory} 

\bigskip

\author{F. T. Brandt$^\dagger$, Ashok Das$^\ddagger$, J. Frenkel$^\dagger$ 
and A. J. da Silva$^\dagger$}
\address{$^\dagger$Instituto de F\'\i sica,
Universidade de S\~ao Paulo\\
S\~ao Paulo, SP 05315-970, BRAZIL\\
$^\ddagger$Department of Physics and Astronomy,
University of Rochester\\
Rochester, NY 14627, USA}

\date{September   1998}
\maketitle 

\medskip

\begin{abstract}
In this  paper, we  give a simple   diagrammatic identification of the
unique combination of the causal $n$-point vertex functions 
in the real time
formalism that would coincide with  the corresponding functions
obtained   in  the imaginary time   formalism. Furthermore,  we give a
simple calculational  method for evaluating the  temperature dependent
parts  of the  retarded   vertex functions, to one  loop,  by
identifying   them  with  the  forward   scattering  amplitudes
of on-shell thermal particles. As an application of the method, 
we  calculate and show that the temperature dependent parts of  
all  the retarded functions  vanish at one loop order
for $1+1$ dimensional massless QED. We further point out that,
in this model, in  fact,  the temperature  dependent parts of  all the
retarded vertex functions vanish to all orders in perturbation theory. 
\end{abstract}

\section{Introduction}\label{sect1}

\par
In recent years, there has been an increased  interest in the study of
finite temperature  field theory primarily from  the point  of view of
understanding the structure   of the  early universe  as  well  as the
properties of the quark gluon  plasma, the latter bearing relevance to
experiments planned  in the near future \cite{gross:1981br,weldon:1982aqweldon:1983jn,kajantie:1985xx,braaten:1990mzbraaten:1990az,smilga:1996cm}.
These  studies, in turn, have
led to a better understanding  of finite temperature field theories in
general and new structures continue to emerge. 

One  of the  important things,  in the  context of  finite temperature
field theories,  is to find  simple  calculational  rules for  various
quantities of physical importance.  Some of  the physical phenomena of
interest at  finite temperature, such as  the linear response, involve
retarded functions  as opposed to the  time ordered functions  that we
are used to in quantum field theories at zero temperature.
As is well known \cite{kapusta:book89,lebellac:book96,das:book97},
 there are two distinct ways of calculating statistical averages
-- commonly known  as  the imaginary time formalism  and  the real time
formalism (furthermore, there are two  real time formalisms).  In  the
imaginary    time  formalism, the calculation     of  the retarded (or
advanced)  functions   is quite   straightforward.   We calculate  the
relevant vertex functions in 
a Euclidean field theory with appropriate (anti)
periodic boundary   conditions  and  then  analytically  continue  the
resulting function  to   the    appropriate   axis  in the     complex   
energy
plane. Furthermore, there exists a very simple method for calculating 
these functions to one loop order, which relates them to 
forward scattering amplitudes of on-shell thermal particles
\cite{frenkel:1991tsbrandt:1997se}.
In contrast, in the real
time formalism, we  have a doubling  of fields which for the $n$-point
functions leads to $(2^{n}-1)$ independent causal functions. It is, of
course, not clear {\it  a priori} whether there  even exists a  unique
definition of a retarded $n$-point function in the real time formalism
and if so, how  it compares with the  retarded function calculated in
the imaginary time  formalism. It is  worth  pointing out here that  a
meaningful definition of retarded vertex functions in quantum field theories
at zero temperature already exists and a lot  of work has been done in
recent times  showing how this  definition, when generalized to finite
temperature, coincides with the quantities obtained from the imaginary
time  formalism\cite{landsman:1987uw,evans:1992kyevans:1993ak,baier:1994yh,carrington:1996rx}.
In this  paper,  we would  like  to
describe a simple calculational method which gives in the real time 
formalism one loop retarded $n$-point function
quite easily. 

In  Sect. \ref{sect2},  we  describe  briefly a  simple approach  
that works well  in the imaginary time  formalism, at least to
one loop. In Sect. \ref{sect3}, we explicitly identify the diagrams that
correspond to the imaginary  time retarded amplitudes and then drawing
from the  results   of Sect. \ref{sect2}, present a  simple method   for
calculating  these amplitudes in  the  real time formalism.
In this approach, the temperature dependent parts of the
retarded $n$-point functions are given a simple diagrammatic
representation, which expresses them in terms of Feynman
amplitudes with physical (retarded/advanced) propagators
and a single statistical factor.
As an  application of
this method, we  show  in Sect. \ref{sect4}
that  all the retarded vertex functions for  a
$1+1$ dimensional fermion interacting with an  external gauge field at
finite temperature  vanish to one loop order. We also 
extract  some  further  properties  of  this model  to   show, in 
Sect. \ref{sect5}, that the
retarded  self-energy, as  well  as all  the  other retarded $n$-point
functions actually vanish to all higher orders.  
We end our paper with a brief conclusion in Sect. \ref{sect6}. 

\section{Imaginary Time Calculational Method}\label{sect2}

The calculation  of an  amplitude  in the imaginary time  formalism is
exactly  similar to that at  zero temperature.  The only difference is
that the energy, instead  of taking continuous values, takes discrete
values depending on  the (anti) periodic  boundary condition used ($T$
is the temperature) 
\begin{equation}
\omega  = \left\{\begin{array}{cl}   {2n\pi\over   T} &  {\rm  for\;\;
                 bosons}\\ {(2n+1)\pi\over T} & {\rm for\;\; fermions}
                 \end{array}\right.  .
\end{equation}  
Consequently, when one has  internal loops, the loop energy  variable,
rather being integrated, is summed over  all possible discrete values.
The sum, at one  loop, gives rise  to a single statistical factor  for
the particle  type whose  energy  is being  summed in the  temperature
dependent part of  the amplitude.  This  temperature dependent part of
the amplitude can  be represented  as  a forward scattering  amplitude
where  the   internal line  is   cut open  to  be   on-shell with  the
appropriate statistical factor\cite{frenkel:1991tsbrandt:1997se}
(and the permutations of  the external
lines which corresponds to different  cuttings of the internal 
lines).
Calculationally, this  is indeed quite  simple. Let us illustrate this
with the  simple example  of the  fermionic contribution to  the  self
energy of an Abelian  gauge boson in  $3+1$ dimensions.  The one  loop
amplitude can be  written in terms  of  two diagrams as it is indicated
in Fig. (1) 
(the momentum of every   internal line that is    being cut can
always be redefined to  be the momentum that is  being integrated 
which leads to the same graph with different permutations of the 
external lines). 
\begin{figure}[h!]
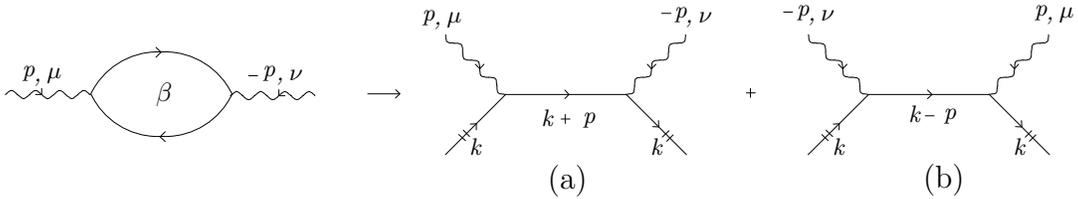

\center{
\input ForwardTwo.eepic
}
\bigskip
   \label{fig1}
\caption{The forward scattering amplitudes associated with the
thermal two-point function. The momentum $k$ of the thermal particles
is on-shell, being integrated over with the statistical
factor $n_{F}(\vert k_0\vert)$}
  \end{figure}

The amplitude itself can now be evaluated trivially to give 
\begin{eqnarray}
\Pi^{\mu\nu\;(\beta)}(p)   &    =   &   -   e^{2}\int    {d^{4}k\over
(2\pi)^{3}}\;n_{F}(|k^{0}|)\delta   (k^{2} -    m^{2})\nonumber\\  & &
\hspace{.3in}\times\left({N^{\mu\nu}(k,p)\over ((k+p)^{2} - m^{2})}  +
{N^{\nu\mu}(k-p,p)\over ((k-p)^{2} - m^{2})}\right). \label{a1} 
\end{eqnarray}
Here $\beta$ is  the inverse  temperature in  units of the  Boltzmann
constant, 
\[
n_{F}(|k^{0}|) = {1\over e^{\beta |k^{0}|} + 1} 
\]
and the tensor structure arising from the fermion loop is given by
\[
N^{\mu\nu}(k,p)   = {\rm     Tr}\left(\gamma^{\mu}(k\!\!\!\slash     +
p\!\!\!\slash + m)\gamma^{\nu} (k\!\!\!\slash + m)\right). 
\]
The high temperature limit of  the  self-energy can now be  
easily calculated and has the form 
\begin{equation}
\Pi^{{\mu\nu}\;(\beta)}\simeq \frac{e^2 T^2}{3}
\int \frac{d\Omega}{4\pi} \left(\frac{p^0\hat k^\mu \hat k^\nu}
{p\cdot\hat k} - \eta^{\mu 0}  \eta^{\nu 0}\right),  
\end{equation}
where the angular integration is performed in the 3-dimensional
space and $\hat k$ is a light-like 4-vector given by 
$\hat k = (1,{\bf\hat k})$.

There  are some aspects to note here.
First, in the above expression, $p^{0}={2ni\pi\over T}$ even though we
have treated   it  as a real    variable.   Second, even  though  this
calculation appears to be similar to a real time calculation, there is
an essential difference, namely, the  propagators in Eq (\ref{a1}) do
not have the $i\epsilon$ term. 

This can, in fact, be generalized to the  $n$-point one loop amplitude
in a straightforward manner. For example, the temperature dependent 
part of the 3-point function has the diagrammatic representation
shown in Fig. (2).

\begin{figure}[h!]
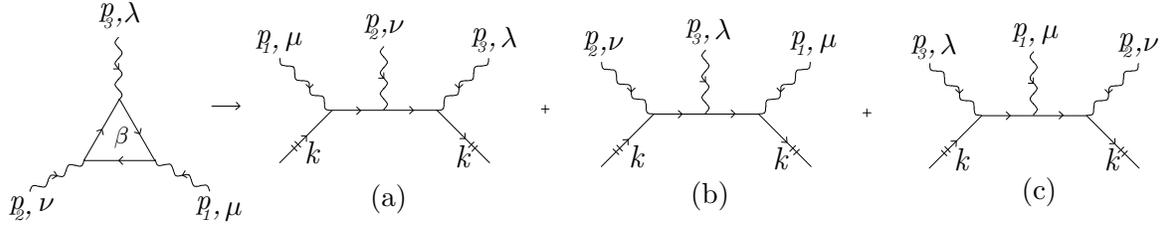

\center{\input ForwardThree.eepic
}
\bigskip
   \label{fig2} 
\caption{The forward scattering amplitudes associated with the
thermal 3-point function. All external momenta $p_i$ are inward with
$\sum_i p_i =0$.}
  \end{figure}

It is important to note that this method also
allows us to  calculate the temperature dependent physical amplitudes,
such  as  the retarded amplitudes,  in  a  simple manner  in terms  of
forward scattering amplitude diagrams with physical propagators. Thus,
the temperature dependent  part of the retarded $2$-point
amplitude can be obtained by analytical continuation of the
external energy $p^0\rightarrow E+i\epsilon$.
In general, the retarded $n$-point amplitude is found by the
analytic continuation of the energy
$p^0_n\rightarrow E_n+(n-1)i\epsilon$
(all other external energies being analytically continued as
$p^0_j\rightarrow E_j-i\epsilon, j=1,2,\cdots ,n-1$),
where we  have assumed that the vertex  with external momentum $p_{n}$
corresponds to the  one   with the largest   time  in the   coordinate
space. Such amplitudes are, again, straightforward to calculate. 
They contain physical (advanced or retarded) propagators
$G_{A,R}$ in the intermediate states, as illustrated in 
Fig. (3).
\begin{figure}[h!]
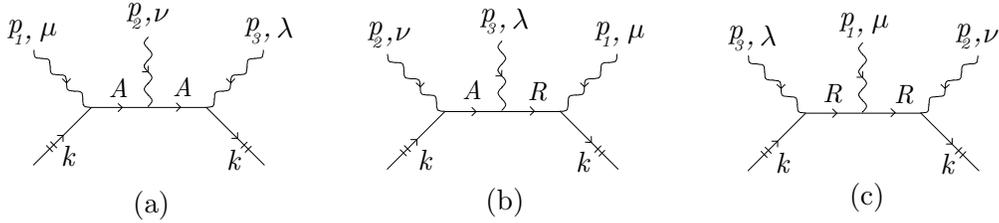

\center{
\input ForwardThreeRA.eepic
}
\bigskip
   \label{fig3a} 
\caption{Examples of forward scattering amplitudes 
with physical [retarded(R) or advanced (A)] propagators.}
  \end{figure}

\section{Real Time Method for Retarded Functions}\label{sect3}

The physical   two  point functions, such   as  the retarded   and the
advanced vertex functions, are  well known in terms of the causal
vertex functions. Thus, for example, ($\Sigma_{+-}$  is defined with a
negative sign) 
\begin{equation}
\Sigma_{R} (p) = \Sigma_{++} (p) + \Sigma_{+-} (p)\label{b1} 
\end{equation}
and so on. The  problem, however, arises when  we  go to higher  point
functions.  Because  of the doubling  of the field degrees of freedom,
there are $(2^{n}-1)$   independent causal $n$-point   amplitudes. The
main difficulty is to determine which combination of these independent
amplitudes will correspond to the  retarded amplitude that is obtained
from  the   calculations in the  imaginary time   formalism and how to
evaluate them in a simple way. In  this connection, it is worth noting
from the discussion  in  the previous  section that  at  the one  loop
order,  the  temperature dependent  amplitude  in  the  imaginary time
formalism involves only a  single statistical factor. In contrast, the
temperature dependent  part  of every  propagator,  in  the real  time
formalism  has a  statistical   factor. Consequently,  every  $n$-point
causal amplitude  can, in  principle, involve up to  a maximum   of $n$
statistical factors.  Thus, the combination  of the  causal amplitudes
that  will  correspond to the    calculations  of the  imaginary  time
formalism has to be such that all the higher order statistical factors
except  the linear terms  cancel out  in the  combination. This is not
hard to determine.   In fact, let us  start  with some explicit  lower
order results before giving the general result. 

Our discussion will  be in the closed  time path formalism even though
everything  can  be  described equally   well  in  the   formalism  of
thermo field dynamics. We note that  in the closed time path formalism,
the propagator has a $2\times 2$ matrix form given by 
\begin{equation}
G_{a b}    (q)     =    G_{a b}^{(0)}    (q)         +
G_{a b}^{(\beta)} (q)\hspace{.3in}a ,b = \pm .
\end{equation}
Here  $G_{a b}^{(0)}(q)$ is  the zero  temperature propagator
whereas $G_{a b}^{(\beta)} (q)$  corresponds  to  the on-shell
thermal correction to   the    propagator.  To keep     the discussion
completely general,  we  do not  take  any  particular  form  for  the
propagator. Thus, $G_{a b}$ can represent the propagator for a
boson or  a  fermion with  the Dirac matrix   structure factored.  The
important thing to note is that no matter what is the propagator being
considered, the temperature dependent part is the  same for all values
of the  indices $a,\,b$. For  simplicity of discussion,  let us
introduce a diagrammatic   representation for the   two parts  of  the
propagator. Thus, a simple line would represent the  zero temperature part of
the  propagator while a line  with a cut  would  represent the thermal
correction: 
\begin{equation}
G_{a b}^{(0)}(q)=
  \begin{array}[h]{c}
\setlength{\unitlength}{0.00083333in}
\begingroup\makeatletter\ifx\SetFigFont\undefined%
\gdef\SetFigFont#1#2#3#4#5{%
  \reset@font\fontsize{#1}{#2pt}%
  \fontfamily{#3}\fontseries{#4}\fontshape{#5}%
  \selectfont}%
\fi\endgroup%
{\renewcommand{\dashlinestretch}{30}
\begin{picture}(1053,369)(0,-10)
\path(70,88)(970,88)
\path(473,148)(538,86)(472,22)
\put(0,171){\makebox(0,0)[lb]{\smash{{{\SetFigFont{12}{14.4}{\rmdefault}{\mddefault}{\itdefault}a}}}}}
\put(961,152){\makebox(0,0)[lb]{\smash{{{\SetFigFont{12}{14.4}{\rmdefault}{\mddefault}{\itdefault}b}}}}}
\put(464,246){\makebox(0,0)[lb]{\smash{{{\SetFigFont{12}{14.4}{\rmdefault}{\mddefault}{\itdefault}q}}}}}
\end{picture}
}

\end{array};
\;\;\;\;\;\;\;
G_{a b}^{(\beta)}(q)=
  \begin{array}[h]{c}
\setlength{\unitlength}{0.00083333in}
\begingroup\makeatletter\ifx\SetFigFont\undefined%
\gdef\SetFigFont#1#2#3#4#5{%
  \reset@font\fontsize{#1}{#2pt}%
  \fontfamily{#3}\fontseries{#4}\fontshape{#5}%
  \selectfont}%
\fi\endgroup%
{\renewcommand{\dashlinestretch}{30}
\begin{picture}(1053,374)(0,-10)
\path(580,22)(580,158)
\path(640,23)(640,157)
\path(367,155)(432,93)(366,29)
\path(70,93)(970,93)
\put(0,176){\makebox(0,0)[lb]{\smash{{{\SetFigFont{12}{14.4}{\rmdefault}{\mddefault}{\itdefault}a}}}}}
\put(961,157){\makebox(0,0)[lb]{\smash{{{\SetFigFont{12}{14.4}{\rmdefault}{\mddefault}{\itdefault}b}}}}}
\put(464,251){\makebox(0,0)[lb]{\smash{{{\SetFigFont{12}{14.4}{\rmdefault}{\mddefault}{\itdefault}q}}}}}
\end{picture}
}
\end{array} = G^{(\beta)}(q) .
\end{equation}
We also note the following general relations\cite{das:book97}:
\begin{equation}\label{propag}
\begin{array}{lllll}
G_R & = & G_{++} - G_{+-} & = & G_{-+} - G_{--} \\
{} & {} & {} & {} & {} \\
G_A & = & G_{++} - G_{-+} & = & G_{+-} - G_{--} 
\end{array},
\end{equation}
which will be useful in what follows.
Furthermore, we remark that in  the closed time path formalism,
there are two  kinds  of vertices, the  vertices  for  the ``$-$''  fields
having a relative negative sign compared to those for the ``$+$'' fields. 

With these observations, let us first look at the retarded 
two-point function in the real time formalism. We then see
that the sum of the graphs with two cut propagators vanishes (to 
keep the discussion general, we will use dashed external lines and solid 
internal lines which can stand for any field):
\begin{equation}
  \begin{array}[h]{c}
\setlength{\unitlength}{0.00036667in}
\begingroup\makeatletter\ifx\SetFigFont\undefined%
\gdef\SetFigFont#1#2#3#4#5{%
  \reset@font\fontsize{#1}{#2pt}%
  \fontfamily{#3}\fontseries{#4}\fontshape{#5}%
  \selectfont}%
\fi\endgroup%
{\renewcommand{\dashlinestretch}{30}
\begin{picture}(3752,1808)(0,-10)
\thicklines
\path(711,778)(711,914)
\path(646,848)(781,848)
\path(3041,778)(3041,914)
\path(3106,848)(2971,848)
\path(1948,22)(1948,157)
\path(2038,22)(2038,157)
\path(1793,1419)(1793,1284)
\path(1703,1419)(1703,1284)
\thinlines
\path(1808,156)(1743,92)(1808,29)
\path(438,665)(496,722)(438,777)
\path(202,720)(292,719)
\path(382,719)(472,720)
\path(562,719)(652,720)
\path(22,719)(112,719)
\path(742,719)(832,719)
\put(1829.878,240.123){\arc{2220.110}{3.5846}{4.7585}}
\put(1827.029,1194.134){\arc{2210.905}{1.5220}{2.7013}}
\path(1808,156)(1743,92)(1808,29)
\path(1933,1285)(1998,1349)(1933,1412)
\path(3314,665)(3256,722)(3314,777)
\path(3550,720)(3460,719)
\path(3370,719)(3280,720)
\path(3190,719)(3100,720)
\path(3730,719)(3640,719)
\path(3010,719)(2920,719)
\put(1922.122,240.123){\arc{2220.111}{4.6663}{5.8401}}
\put(1924.971,1194.134){\arc{2210.905}{0.4403}{1.6196}}
\path(3383,954)(3248,954)
\put(333,905){\makebox(0,0)[lb]{\smash{{{\SetFigFont{10}{12.0}{\rmdefault}{\mddefault}{\itdefault}p}}}}}
\put(3426,900){\makebox(0,0)[lb]{\smash{{{\SetFigFont{10}{12.0}{\rmdefault}{\mddefault}{\itdefault}p}}}}}
\end{picture}
}
\end{array}\;\;\; +\;\;\;
  \begin{array}[h]{c}
\setlength{\unitlength}{0.00036667in}
\begingroup\makeatletter\ifx\SetFigFont\undefined%
\gdef\SetFigFont#1#2#3#4#5{%
  \reset@font\fontsize{#1}{#2pt}%
  \fontfamily{#3}\fontseries{#4}\fontshape{#5}%
  \selectfont}%
\fi\endgroup%
{\renewcommand{\dashlinestretch}{30}
\begin{picture}(3752,1808)(0,-10)
\thicklines
\path(711,778)(711,914)
\path(646,848)(781,848)
\path(3106,848)(2971,848)
\path(1948,22)(1948,157)
\path(2038,22)(2038,157)
\path(1793,1419)(1793,1284)
\path(1703,1419)(1703,1284)
\thinlines
\path(1808,156)(1743,92)(1808,29)
\path(438,665)(496,722)(438,777)
\path(202,720)(292,719)
\path(382,719)(472,720)
\path(562,719)(652,720)
\path(22,719)(112,719)
\path(742,719)(832,719)
\put(1829.878,240.123){\arc{2220.110}{3.5846}{4.7585}}
\put(1827.029,1194.134){\arc{2210.905}{1.5220}{2.7013}}
\path(1808,156)(1743,92)(1808,29)
\path(1933,1285)(1998,1349)(1933,1412)
\path(3314,665)(3256,722)(3314,777)
\path(3550,720)(3460,719)
\path(3370,719)(3280,720)
\path(3190,719)(3100,720)
\path(3730,719)(3640,719)
\path(3010,719)(2920,719)
\put(1922.122,240.123){\arc{2220.111}{4.6663}{5.8401}}
\put(1924.971,1194.134){\arc{2210.905}{0.4403}{1.6196}}
\path(3383,954)(3248,954)
\put(333,905){\makebox(0,0)[lb]{\smash{{{\SetFigFont{10}{12.0}{\rmdefault}{\mddefault}{\itdefault}p}}}}}
\put(3426,900){\makebox(0,0)[lb]{\smash{{{\SetFigFont{10}{12.0}{\rmdefault}{\mddefault}{\itdefault}p}}}}}
\end{picture}
}
\end{array}\;\;\; =\;\;\; 0 .
\end{equation}

Next, using the relations (\ref{propag}),  we obtain the  
following diagrammatic equation for the  sum of 
diagrams  with a single cut line:
\begin{equation}\label{sigma1cut}
  \begin{array}[h]{c}
\setlength{\unitlength}{0.00036667in}
\begingroup\makeatletter\ifx\SetFigFont\undefined%
\gdef\SetFigFont#1#2#3#4#5{%
  \reset@font\fontsize{#1}{#2pt}%
  \fontfamily{#3}\fontseries{#4}\fontshape{#5}%
  \selectfont}%
\fi\endgroup%
{\renewcommand{\dashlinestretch}{30}
\begin{picture}(3752,1808)(0,-10)
\thicklines
\path(711,778)(711,914)
\path(646,848)(781,848)
\path(3041,778)(3041,914)
\path(3106,848)(2971,848)
\path(1948,22)(1948,157)
\path(2038,22)(2038,157)
\thinlines
\path(1808,156)(1743,92)(1808,29)
\path(438,665)(496,722)(438,777)
\path(202,720)(292,719)
\path(382,719)(472,720)
\path(562,719)(652,720)
\path(22,719)(112,719)
\path(742,719)(832,719)
\put(1829.878,240.123){\arc{2220.110}{3.5846}{4.7585}}
\put(1827.029,1194.134){\arc{2210.905}{1.5220}{2.7013}}
\path(1808,156)(1743,92)(1808,29)
\path(1933,1285)(1998,1349)(1933,1412)
\path(3314,665)(3256,722)(3314,777)
\path(3550,720)(3460,719)
\path(3370,719)(3280,720)
\path(3190,719)(3100,720)
\path(3730,719)(3640,719)
\path(3010,719)(2920,719)
\put(1922.122,240.123){\arc{2220.111}{4.6663}{5.8401}}
\put(1924.971,1194.134){\arc{2210.905}{0.4403}{1.6196}}
\path(3383,954)(3248,954)
\put(333,905){\makebox(0,0)[lb]{\smash{{{\SetFigFont{10}{12.0}{\rmdefault}{\mddefault}{\itdefault}p}}}}}
\put(3426,900){\makebox(0,0)[lb]{\smash{{{\SetFigFont{10}{12.0}{\rmdefault}{\mddefault}{\itdefault}p}}}}}
\put(1800,-300){\makebox(0,0)[lb]{\smash{{{\SetFigFont{10}{12.0}{\rmdefault}{\mddefault}{\itdefault}k}}}}}
\end{picture}
}
\end{array}\; +\;
  \begin{array}[h]{c}
\setlength{\unitlength}{0.00036667in}
\begingroup\makeatletter\ifx\SetFigFont\undefined%
\gdef\SetFigFont#1#2#3#4#5{%
  \reset@font\fontsize{#1}{#2pt}%
  \fontfamily{#3}\fontseries{#4}\fontshape{#5}%
  \selectfont}%
\fi\endgroup%
{\renewcommand{\dashlinestretch}{30}
\begin{picture}(3752,1808)(0,-10)
\thicklines
\path(711,778)(711,914)
\path(646,848)(781,848)
\path(3106,848)(2971,848)
\path(1948,22)(1948,157)
\path(2038,22)(2038,157)
\thinlines
\path(1808,156)(1743,92)(1808,29)
\path(438,665)(496,722)(438,777)
\path(202,720)(292,719)
\path(382,719)(472,720)
\path(562,719)(652,720)
\path(22,719)(112,719)
\path(742,719)(832,719)
\put(1829.878,240.123){\arc{2220.110}{3.5846}{4.7585}}
\put(1827.029,1194.134){\arc{2210.905}{1.5220}{2.7013}}
\path(1808,156)(1743,92)(1808,29)
\path(1933,1285)(1998,1349)(1933,1412)
\path(3314,665)(3256,722)(3314,777)
\path(3550,720)(3460,719)
\path(3370,719)(3280,720)
\path(3190,719)(3100,720)
\path(3730,719)(3640,719)
\path(3010,719)(2920,719)
\put(1922.122,240.123){\arc{2220.111}{4.6663}{5.8401}}
\put(1924.971,1194.134){\arc{2210.905}{0.4403}{1.6196}}
\path(3383,954)(3248,954)
\put(333,905){\makebox(0,0)[lb]{\smash{{{\SetFigFont{10}{12.0}{\rmdefault}{\mddefault}{\itdefault}p}}}}}
\put(3426,900){\makebox(0,0)[lb]{\smash{{{\SetFigFont{10}{12.0}{\rmdefault}{\mddefault}{\itdefault}p}}}}}
\put(1800,-300){\makebox(0,0)[lb]{\smash{{{\SetFigFont{10}{12.0}{\rmdefault}{\mddefault}{\itdefault}k}}}}}
\end{picture}
}
\end{array}\; =\;
  \begin{array}[h]{c}
\setlength{\unitlength}{0.00036667in}
\begingroup\makeatletter\ifx\SetFigFont\undefined%
\gdef\SetFigFont#1#2#3#4#5{%
  \reset@font\fontsize{#1}{#2pt}%
  \fontfamily{#3}\fontseries{#4}\fontshape{#5}%
  \selectfont}%
\fi\endgroup%
{\renewcommand{\dashlinestretch}{30}
\begin{picture}(3752,1808)(0,-10)
\thicklines
\path(1948,22)(1948,157)
\path(2038,22)(2038,157)
\thinlines
\path(1808,156)(1743,92)(1808,29)
\path(438,665)(496,722)(438,777)
\path(202,720)(292,719)
\path(382,719)(472,720)
\path(562,719)(652,720)
\path(22,719)(112,719)
\path(742,719)(832,719)
\put(1829.878,240.123){\arc{2220.110}{3.5846}{4.7585}}
\put(1827.029,1194.134){\arc{2210.905}{1.5220}{2.7013}}
\path(1808,156)(1743,92)(1808,29)
\path(1933,1285)(1998,1349)(1933,1412)
\path(3314,665)(3256,722)(3314,777)
\path(3550,720)(3460,719)
\path(3370,719)(3280,720)
\path(3190,719)(3100,720)
\path(3730,719)(3640,719)
\path(3010,719)(2920,719)
\put(1922.122,240.123){\arc{2220.111}{4.6663}{5.8401}}
\put(1924.971,1194.134){\arc{2210.905}{0.4403}{1.6196}}
\path(3383,954)(3248,954)
\put(333,905){\makebox(0,0)[lb]{\smash{{{\SetFigFont{10}{12.0}{\rmdefault}{\mddefault}{\itdefault}p}}}}}
\put(1777,1598){\makebox(0,0)[lb]{\smash{{{\SetFigFont{9}{10.8}{\rmdefault}{\mddefault}{\itdefault}R}}}}}
\put(3426,900){\makebox(0,0)[lb]{\smash{{{\SetFigFont{10}{12.0}{\rmdefault}{\mddefault}{\itdefault}p}}}}}
\put(1800,-300){\makebox(0,0)[lb]{\smash{{{\SetFigFont{10}{12.0}{\rmdefault}{\mddefault}{\itdefault}k}}}}}
\end{picture}
} .
\end{array}
\end{equation}
The graph on the right-hand side of (\ref{sigma1cut}) coincides
with the forward scattering amplitude in Fig. (1a)
obtained in the imaginary time formalism,
by analytic continuation of $p_0\rightarrow E+i\epsilon$.
Similarly, the sum of the diagrams where only the other propagator is
cut, corresponds to the amplitude
shown in Fig. (1b) (these amplitudes actually give
 the same contribution).

Let us consider next the sum of the 
three point functions in the 
real-time formalism. It is clear that the sum of the following graphs with 
three cut propagators vanishes:
\begin{equation}\label{three3cuts}
  \begin{array}[h]{c}
\setlength{\unitlength}{0.00033333in}
\begingroup\makeatletter\ifx\SetFigFont\undefined%
\gdef\SetFigFont#1#2#3#4#5{%
  \reset@font\fontsize{#1}{#2pt}%
  \fontfamily{#3}\fontseries{#4}\fontshape{#5}%
  \selectfont}%
\fi\endgroup%
{\renewcommand{\dashlinestretch}{30}
\begin{picture}(3328,2820)(0,-10)
\thicklines
\path(1698,357)(1698,493)
\path(1788,356)(1788,493)
\path(2146,1152)(2029,1084)
\path(2101,1230)(1984,1162)
\path(1217,1036)(1100,1104)
\path(1172,958)(1055,1026)
\path(646,460)(646,596)
\path(581,530)(716,530)
\path(2589,490)(2589,626)
\path(2524,560)(2659,560)
\path(1876,1760)(1876,1896)
\path(1811,1830)(1946,1830)
\thinlines
\path(1691,2290)(1634,2232)(1579,2290)
\path(1636,2526)(1637,2436)
\path(1637,2346)(1636,2256)
\path(1637,2166)(1636,2076)
\path(1637,2706)(1637,2616)
\path(1637,1986)(1637,1896)
\path(414,277)(494,257)(470,180)
\path(238,111)(315,158)
\path(393,203)(472,246)
\path(549,293)(628,336)
\path(81,22)(159,68)
\path(705,383)(783,427)
\path(2100,964)(2188,940)(2211,1027)
\path(1557,492)(1492,428)(1558,363)
\path(2799,183)(2779,261)(2855,280)
\path(3031,113)(2953,157)
\path(2875,202)(2797,249)
\path(2719,292)(2641,339)
\path(3188,22)(3109,67)
\path(2563,382)(2485,427)
\path(785,425)(2487,425)
\path(785,425)(1636,1900)
\path(1636,1901)(2487,426)
\path(1171,1224)(1259,1248)(1282,1161)  
\put(1745,2649){\makebox(0,0)[lb]{\smash{{{\SetFigFont{8}{9.6}{\rmdefault}{\mddefault}{\itdefault}p}}}}}
\put(1880,2539){\makebox(0,0)[lb]{\smash{{{\SetFigFont{5}{6.0}{\rmdefault}{\mddefault}{\itdefault}3}}}}}
\put(0,292){\makebox(0,0)[lb]{\smash{{{\SetFigFont{8}{9.6}{\rmdefault}{\mddefault}{\itdefault}p}}}}}
\put(3097,292){\makebox(0,0)[lb]{\smash{{{\SetFigFont{8}{9.6}{\rmdefault}{\mddefault}{\itdefault}p}}}}}
\put(130,187){\makebox(0,0)[lb]{\smash{{{\SetFigFont{5}{6.0}{\rmdefault}{\mddefault}{\itdefault}2}}}}}
\put(3252,197){\makebox(0,0)[lb]{\smash{{{\SetFigFont{5}{6.0}{\rmdefault}{\mddefault}{\itdefault}1}}}}}
\end{picture}
} 
\end{array}\; +\; 
  \begin{array}[h]{c}
\setlength{\unitlength}{0.00033333in}
\begingroup\makeatletter\ifx\SetFigFont\undefined%
\gdef\SetFigFont#1#2#3#4#5{%
  \reset@font\fontsize{#1}{#2pt}%
  \fontfamily{#3}\fontseries{#4}\fontshape{#5}%
  \selectfont}%
\fi\endgroup%
{\renewcommand{\dashlinestretch}{30}
\begin{picture}(3328,2820)(0,-10)
\thicklines
\path(1698,357)(1698,493)
\path(1788,356)(1788,493)
\path(2146,1152)(2029,1084)
\path(2101,1230)(1984,1162)
\path(1217,1036)(1100,1104)
\path(1172,958)(1055,1026)
\path(646,460)(646,596)
\path(581,530)(716,530)
\path(2524,560)(2659,560)
\path(1876,1760)(1876,1896)
\path(1811,1830)(1946,1830)
\thinlines
\path(1691,2290)(1634,2232)(1579,2290)
\path(1636,2526)(1637,2436)
\path(1637,2346)(1636,2256)
\path(1637,2166)(1636,2076)
\path(1637,2706)(1637,2616)
\path(1637,1986)(1637,1896)
\path(414,277)(494,257)(470,180)
\path(238,111)(315,158)
\path(393,203)(472,246)
\path(549,293)(628,336)
\path(81,22)(159,68)
\path(705,383)(783,427)
\path(2100,964)(2188,940)(2211,1027)
\path(1557,492)(1492,428)(1558,363)
\path(2799,183)(2779,261)(2855,280)
\path(3031,113)(2953,157)
\path(2875,202)(2797,249)
\path(2719,292)(2641,339)
\path(3188,22)(3109,67)
\path(2563,382)(2485,427)
\path(785,425)(2487,425)
\path(785,425)(1636,1900)
\path(1636,1901)(2487,426)
\path(1299,1190)(1276,1279)(1190,1254)
\put(1745,2649){\makebox(0,0)[lb]{\smash{{{\SetFigFont{8}{9.6}{\rmdefault}{\mddefault}{\itdefault}p}}}}}
\put(1880,2539){\makebox(0,0)[lb]{\smash{{{\SetFigFont{5}{6.0}{\rmdefault}{\mddefault}{\itdefault}3}}}}}
\put(0,292){\makebox(0,0)[lb]{\smash{{{\SetFigFont{8}{9.6}{\rmdefault}{\mddefault}{\itdefault}p}}}}}
\put(3097,292){\makebox(0,0)[lb]{\smash{{{\SetFigFont{8}{9.6}{\rmdefault}{\mddefault}{\itdefault}p}}}}}
\put(130,187){\makebox(0,0)[lb]{\smash{{{\SetFigFont{5}{6.0}{\rmdefault}{\mddefault}{\itdefault}2}}}}}
\put(3252,197){\makebox(0,0)[lb]{\smash{{{\SetFigFont{5}{6.0}{\rmdefault}{\mddefault}{\itdefault}1}}}}}
\end{picture}
}
\end{array}\; +\;  
  \begin{array}[h]{c}
\setlength{\unitlength}{0.00033333in}
\begingroup\makeatletter\ifx\SetFigFont\undefined%
\gdef\SetFigFont#1#2#3#4#5{%
  \reset@font\fontsize{#1}{#2pt}%
  \fontfamily{#3}\fontseries{#4}\fontshape{#5}%
  \selectfont}%
\fi\endgroup%
{\renewcommand{\dashlinestretch}{30}
\begin{picture}(3328,2820)(0,-10)
\thicklines
\path(1698,357)(1698,493)
\path(1788,356)(1788,493)
\path(2146,1152)(2029,1084)
\path(2101,1230)(1984,1162)
\path(1217,1036)(1100,1104)
\path(1172,958)(1055,1026)
\path(581,530)(716,530)
\path(2589,490)(2589,626)
\path(2524,560)(2659,560)
\path(1876,1760)(1876,1896)
\path(1811,1830)(1946,1830)
\thinlines
\path(1691,2290)(1634,2232)(1579,2290)
\path(1636,2526)(1637,2436)
\path(1637,2346)(1636,2256)
\path(1637,2166)(1636,2076)
\path(1637,2706)(1637,2616)
\path(1637,1986)(1637,1896)
\path(414,277)(494,257)(470,180)
\path(238,111)(315,158)
\path(393,203)(472,246)
\path(549,293)(628,336)
\path(81,22)(159,68)
\path(705,383)(783,427)
\path(2100,964)(2188,940)(2211,1027)
\path(1557,492)(1492,428)(1558,363)
\path(2799,183)(2779,261)(2855,280)
\path(3031,113)(2953,157)
\path(2875,202)(2797,249)
\path(2719,292)(2641,339)
\path(3188,22)(3109,67)
\path(2563,382)(2485,427)
\path(785,425)(2487,425)
\path(785,425)(1636,1900)
\path(1636,1901)(2487,426)
\path(1299,1190)(1276,1279)(1190,1254)
\put(1745,2649){\makebox(0,0)[lb]{\smash{{{\SetFigFont{8}{9.6}{\rmdefault}{\mddefault}{\itdefault}p}}}}}
\put(1880,2539){\makebox(0,0)[lb]{\smash{{{\SetFigFont{5}{6.0}{\rmdefault}{\mddefault}{\itdefault}3}}}}}
\put(0,292){\makebox(0,0)[lb]{\smash{{{\SetFigFont{8}{9.6}{\rmdefault}{\mddefault}{\itdefault}p}}}}}
\put(3097,292){\makebox(0,0)[lb]{\smash{{{\SetFigFont{8}{9.6}{\rmdefault}{\mddefault}{\itdefault}p}}}}}
\put(130,187){\makebox(0,0)[lb]{\smash{{{\SetFigFont{5}{6.0}{\rmdefault}{\mddefault}{\itdefault}2}}}}}
\put(3252,197){\makebox(0,0)[lb]{\smash{{{\SetFigFont{5}{6.0}{\rmdefault}{\mddefault}{\itdefault}1}}}}}
\end{picture}
}
\end{array}\; +\; 
  \begin{array}[h]{c}
\setlength{\unitlength}{0.00033333in}
\begingroup\makeatletter\ifx\SetFigFont\undefined%
\gdef\SetFigFont#1#2#3#4#5{%
  \reset@font\fontsize{#1}{#2pt}%
  \fontfamily{#3}\fontseries{#4}\fontshape{#5}%
  \selectfont}%
\fi\endgroup%
{\renewcommand{\dashlinestretch}{30}
\begin{picture}(3328,2820)(0,-10)
\thicklines
\path(1698,357)(1698,493)
\path(1788,356)(1788,493)
\path(2146,1152)(2029,1084)
\path(2101,1230)(1984,1162)
\path(1217,1036)(1100,1104)
\path(1172,958)(1055,1026)
\path(581,530)(716,530)
\path(2524,560)(2659,560)
\path(1876,1760)(1876,1896)
\path(1811,1830)(1946,1830)
\thinlines
\path(1691,2290)(1634,2232)(1579,2290)
\path(1636,2526)(1637,2436)
\path(1637,2346)(1636,2256)
\path(1637,2166)(1636,2076)
\path(1637,2706)(1637,2616)
\path(1637,1986)(1637,1896)
\path(414,277)(494,257)(470,180)
\path(238,111)(315,158)
\path(393,203)(472,246)
\path(549,293)(628,336)
\path(81,22)(159,68)
\path(705,383)(783,427)
\path(2100,964)(2188,940)(2211,1027)
\path(1557,492)(1492,428)(1558,363)
\path(2799,183)(2779,261)(2855,280)
\path(3031,113)(2953,157)
\path(2875,202)(2797,249)
\path(2719,292)(2641,339)
\path(3188,22)(3109,67)
\path(2563,382)(2485,427)
\path(785,425)(2487,425)
\path(785,425)(1636,1900)
\path(1636,1901)(2487,426)
\path(1299,1190)(1276,1279)(1190,1254)
\put(1745,2649){\makebox(0,0)[lb]{\smash{{{\SetFigFont{8}{9.6}{\rmdefault}{\mddefault}{\itdefault}p}}}}}
\put(1880,2539){\makebox(0,0)[lb]{\smash{{{\SetFigFont{5}{6.0}{\rmdefault}{\mddefault}{\itdefault}3}}}}}
\put(0,292){\makebox(0,0)[lb]{\smash{{{\SetFigFont{8}{9.6}{\rmdefault}{\mddefault}{\itdefault}p}}}}}
\put(3097,292){\makebox(0,0)[lb]{\smash{{{\SetFigFont{8}{9.6}{\rmdefault}{\mddefault}{\itdefault}p}}}}}
\put(130,187){\makebox(0,0)[lb]{\smash{{{\SetFigFont{5}{6.0}{\rmdefault}{\mddefault}{\itdefault}2}}}}}
\put(3252,197){\makebox(0,0)[lb]{\smash{{{\SetFigFont{5}{6.0}{\rmdefault}{\mddefault}{\itdefault}1}}}}}
\end{picture}
}
\end{array}= 0 .
\end{equation}
The diagrams cancel  pairwise    and  this follows simply    from  the
observation that  a cut propagator is the  same no matter what are the
indices $a,b$ as well as the fact that the ``$-$'' vertices have
a relative negative sign compared to the ``$+$'' vertices.  
For the same reason, the sum of the diagrams with two cut propagators
also vanishes. The sum of the
diagrams with  a single  cut propagator, however, do not vanish and,
therefore, it is clear that the sum of this set  of diagrams is likely
to correspond to the temperature dependent  part of the retarded three
point  amplitude obtained  from  the  imaginary time
formalism (because it  will be linear in  the statistical  factor). In
fact, let us  look at  the sum of  these  four diagrams with a   fixed
single cut propagator. Remembering  that a cut propagator is  on-shell
and hence can be thought of as a pair of open lines (on-shell and with
a statistical factor), we can write 
\begin{eqnarray}\label{three1cut}
  \begin{array}[h]{c}
\setlength{\unitlength}{0.00033333in}
\begingroup\makeatletter\ifx\SetFigFont\undefined%
\gdef\SetFigFont#1#2#3#4#5{%
  \reset@font\fontsize{#1}{#2pt}%
  \fontfamily{#3}\fontseries{#4}\fontshape{#5}%
  \selectfont}%
\fi\endgroup%
{\renewcommand{\dashlinestretch}{30}
\begin{picture}(3328,2820)(0,-10)
\thicklines
\path(1698,357)(1698,493)
\path(1788,356)(1788,493)
\path(646,460)(646,596)
\path(581,530)(716,530)
\path(2589,490)(2589,626)
\path(2524,560)(2659,560)
\path(1876,1760)(1876,1896)
\path(1811,1830)(1946,1830)
\thinlines
\path(1691,2290)(1634,2232)(1579,2290)
\path(1636,2526)(1637,2436)
\path(1637,2346)(1636,2256)
\path(1637,2166)(1636,2076)
\path(1637,2706)(1637,2616)
\path(1637,1986)(1637,1896)
\path(414,277)(494,257)(470,180)
\path(238,111)(315,158)
\path(393,203)(472,246)
\path(549,293)(628,336)
\path(81,22)(159,68)
\path(705,383)(783,427)
\path(2100,964)(2188,940)(2211,1027)
\path(1557,492)(1492,428)(1558,363)
\path(2799,183)(2779,261)(2855,280)
\path(3031,113)(2953,157)
\path(2875,202)(2797,249)
\path(2719,292)(2641,339)
\path(3188,22)(3109,67)
\path(2563,382)(2485,427)
\path(785,425)(2487,425)
\path(785,425)(1636,1900)
\path(1636,1901)(2487,426)
\path(1299,1190)(1276,1279)(1190,1254)
\put(1745,2649){\makebox(0,0)[lb]{\smash{{{\SetFigFont{8}{9.6}{\rmdefault}{\mddefault}{\itdefault}p}}}}}
\put(1880,2539){\makebox(0,0)[lb]{\smash{{{\SetFigFont{5}{6.0}{\rmdefault}{\mddefault}{\itdefault}3}}}}}
\put(0,292){\makebox(0,0)[lb]{\smash{{{\SetFigFont{8}{9.6}{\rmdefault}{\mddefault}{\itdefault}p}}}}}
\put(3097,292){\makebox(0,0)[lb]{\smash{{{\SetFigFont{8}{9.6}{\rmdefault}{\mddefault}{\itdefault}p}}}}}
\put(130,187){\makebox(0,0)[lb]{\smash{{{\SetFigFont{5}{6.0}{\rmdefault}{\mddefault}{\itdefault}2}}}}}
\put(3252,197){\makebox(0,0)[lb]{\smash{{{\SetFigFont{5}{6.0}{\rmdefault}{\mddefault}{\itdefault}1}}}}}
\put(1550,50){\makebox(0,0)[lb]{\smash{{{\SetFigFont{8}{9.6}{\rmdefault}{\mddefault}{\itdefault}k}}}}}
\put(1400,-500){\makebox(0,0)[lb]{\smash{{{\SetFigFont{8}{9.6}{\rmdefault}{\mddefault}{\rmdefault}(a)}}}}}
\end{picture}
}
\\
\end{array}\; +\; 
&
  \begin{array}[h]{c}
\setlength{\unitlength}{0.00033333in}
\begingroup\makeatletter\ifx\SetFigFont\undefined%
\gdef\SetFigFont#1#2#3#4#5{%
  \reset@font\fontsize{#1}{#2pt}%
  \fontfamily{#3}\fontseries{#4}\fontshape{#5}%
  \selectfont}%
\fi\endgroup%
{\renewcommand{\dashlinestretch}{30}
\begin{picture}(3328,2820)(0,-10)
\thicklines
\path(1698,357)(1698,493)
\path(1788,356)(1788,493)
\path(646,460)(646,596)
\path(581,530)(716,530)
\path(2524,560)(2659,560)
\path(1876,1760)(1876,1896)
\path(1811,1830)(1946,1830)
\thinlines
\path(1691,2290)(1634,2232)(1579,2290)
\path(1636,2526)(1637,2436)
\path(1637,2346)(1636,2256)
\path(1637,2166)(1636,2076)
\path(1637,2706)(1637,2616)
\path(1637,1986)(1637,1896)
\path(414,277)(494,257)(470,180)
\path(238,111)(315,158)
\path(393,203)(472,246)
\path(549,293)(628,336)
\path(81,22)(159,68)
\path(705,383)(783,427)
\path(2100,964)(2188,940)(2211,1027)
\path(1557,492)(1492,428)(1558,363)
\path(2799,183)(2779,261)(2855,280)
\path(3031,113)(2953,157)
\path(2875,202)(2797,249)
\path(2719,292)(2641,339)
\path(3188,22)(3109,67)
\path(2563,382)(2485,427)
\path(785,425)(2487,425)
\path(785,425)(1636,1900)
\path(1636,1901)(2487,426)
\path(1299,1190)(1276,1279)(1190,1254)
\put(1745,2649){\makebox(0,0)[lb]{\smash{{{\SetFigFont{8}{9.6}{\rmdefault}{\mddefault}{\itdefault}p}}}}}
\put(1880,2539){\makebox(0,0)[lb]{\smash{{{\SetFigFont{5}{6.0}{\rmdefault}{\mddefault}{\itdefault}3}}}}}
\put(0,292){\makebox(0,0)[lb]{\smash{{{\SetFigFont{8}{9.6}{\rmdefault}{\mddefault}{\itdefault}p}}}}}
\put(3097,292){\makebox(0,0)[lb]{\smash{{{\SetFigFont{8}{9.6}{\rmdefault}{\mddefault}{\itdefault}p}}}}}
\put(130,187){\makebox(0,0)[lb]{\smash{{{\SetFigFont{5}{6.0}{\rmdefault}{\mddefault}{\itdefault}2}}}}}
\put(3252,197){\makebox(0,0)[lb]{\smash{{{\SetFigFont{5}{6.0}{\rmdefault}{\mddefault}{\itdefault}1}}}}}
\put(1550,50){\makebox(0,0)[lb]{\smash{{{\SetFigFont{8}{9.6}{\rmdefault}{\mddefault}{\itdefault}k}}}}}
\put(1400,-500){\makebox(0,0)[lb]{\smash{{{\SetFigFont{8}{9.6}{\rmdefault}{\mddefault}{\rmdefault}(b)}}}}}
\end{picture}
}
\\
\end{array}\; +\;  
& {} \nonumber \\  {} & {} \nonumber \\ 
  \begin{array}[h]{c}
\setlength{\unitlength}{0.00033333in}
\begingroup\makeatletter\ifx\SetFigFont\undefined%
\gdef\SetFigFont#1#2#3#4#5{%
  \reset@font\fontsize{#1}{#2pt}%
  \fontfamily{#3}\fontseries{#4}\fontshape{#5}%
  \selectfont}%
\fi\endgroup%
{\renewcommand{\dashlinestretch}{30}
\begin{picture}(3328,2820)(0,-10)
\thicklines
\path(1698,357)(1698,493)
\path(1788,356)(1788,493)
\path(581,530)(716,530)
\path(2589,490)(2589,626)
\path(2524,560)(2659,560)
\path(1876,1760)(1876,1896)
\path(1811,1830)(1946,1830)
\thinlines
\path(1691,2290)(1634,2232)(1579,2290)
\path(1636,2526)(1637,2436)
\path(1637,2346)(1636,2256)
\path(1637,2166)(1636,2076)
\path(1637,2706)(1637,2616)
\path(1637,1986)(1637,1896)
\path(414,277)(494,257)(470,180)
\path(238,111)(315,158)
\path(393,203)(472,246)
\path(549,293)(628,336)
\path(81,22)(159,68)
\path(705,383)(783,427)
\path(2100,964)(2188,940)(2211,1027)
\path(1557,492)(1492,428)(1558,363)
\path(2799,183)(2779,261)(2855,280)
\path(3031,113)(2953,157)
\path(2875,202)(2797,249)
\path(2719,292)(2641,339)
\path(3188,22)(3109,67)
\path(2563,382)(2485,427)
\path(785,425)(2487,425)
\path(785,425)(1636,1900)
\path(1636,1901)(2487,426)
\path(1299,1190)(1276,1279)(1190,1254)
\put(1745,2649){\makebox(0,0)[lb]{\smash{{{\SetFigFont{8}{9.6}{\rmdefault}{\mddefault}{\itdefault}p}}}}}
\put(1880,2539){\makebox(0,0)[lb]{\smash{{{\SetFigFont{5}{6.0}{\rmdefault}{\mddefault}{\itdefault}3}}}}}
\put(0,292){\makebox(0,0)[lb]{\smash{{{\SetFigFont{8}{9.6}{\rmdefault}{\mddefault}{\itdefault}p}}}}}
\put(3097,292){\makebox(0,0)[lb]{\smash{{{\SetFigFont{8}{9.6}{\rmdefault}{\mddefault}{\itdefault}p}}}}}
\put(130,187){\makebox(0,0)[lb]{\smash{{{\SetFigFont{5}{6.0}{\rmdefault}{\mddefault}{\itdefault}2}}}}}
\put(3252,197){\makebox(0,0)[lb]{\smash{{{\SetFigFont{5}{6.0}{\rmdefault}{\mddefault}{\itdefault}1}}}}}
\put(1550,50){\makebox(0,0)[lb]{\smash{{{\SetFigFont{8}{9.6}{\rmdefault}{\mddefault}{\itdefault}k}}}}}
\put(1400,-500){\makebox(0,0)[lb]{\smash{{{\SetFigFont{8}{9.6}{\rmdefault}{\mddefault}{\rmdefault}(c)}}}}}
\end{picture}
}
\\
\end{array}\; +\; 
&
  \begin{array}[h]{c}
\setlength{\unitlength}{0.00033333in}
\begingroup\makeatletter\ifx\SetFigFont\undefined%
\gdef\SetFigFont#1#2#3#4#5{%
  \reset@font\fontsize{#1}{#2pt}%
  \fontfamily{#3}\fontseries{#4}\fontshape{#5}%
  \selectfont}%
\fi\endgroup%
{\renewcommand{\dashlinestretch}{30}
\begin{picture}(3328,2820)(0,-10)
\thicklines
\path(1698,357)(1698,493)
\path(1788,356)(1788,493)
\path(581,530)(716,530)
\path(2524,560)(2659,560)
\path(1876,1760)(1876,1896)
\path(1811,1830)(1946,1830)
\thinlines
\path(1691,2290)(1634,2232)(1579,2290)
\path(1636,2526)(1637,2436)
\path(1637,2346)(1636,2256)
\path(1637,2166)(1636,2076)
\path(1637,2706)(1637,2616)
\path(1637,1986)(1637,1896)
\path(414,277)(494,257)(470,180)
\path(238,111)(315,158)
\path(393,203)(472,246)
\path(549,293)(628,336)
\path(81,22)(159,68)
\path(705,383)(783,427)
\path(2100,964)(2188,940)(2211,1027)
\path(1557,492)(1492,428)(1558,363)
\path(2799,183)(2779,261)(2855,280)
\path(3031,113)(2953,157)
\path(2875,202)(2797,249)
\path(2719,292)(2641,339)
\path(3188,22)(3109,67)
\path(2563,382)(2485,427)
\path(785,425)(2487,425)
\path(785,425)(1636,1900)
\path(1636,1901)(2487,426)
\path(1299,1190)(1276,1279)(1190,1254)
\put(1745,2649){\makebox(0,0)[lb]{\smash{{{\SetFigFont{8}{9.6}{\rmdefault}{\mddefault}{\itdefault}p}}}}}
\put(1880,2539){\makebox(0,0)[lb]{\smash{{{\SetFigFont{5}{6.0}{\rmdefault}{\mddefault}{\itdefault}3}}}}}
\put(0,292){\makebox(0,0)[lb]{\smash{{{\SetFigFont{8}{9.6}{\rmdefault}{\mddefault}{\itdefault}p}}}}}
\put(3097,292){\makebox(0,0)[lb]{\smash{{{\SetFigFont{8}{9.6}{\rmdefault}{\mddefault}{\itdefault}p}}}}}
\put(130,187){\makebox(0,0)[lb]{\smash{{{\SetFigFont{5}{6.0}{\rmdefault}{\mddefault}{\itdefault}2}}}}}
\put(3252,197){\makebox(0,0)[lb]{\smash{{{\SetFigFont{5}{6.0}{\rmdefault}{\mddefault}{\itdefault}1}}}}}
\put(1550,50){\makebox(0,0)[lb]{\smash{{{\SetFigFont{8}{9.6}{\rmdefault}{\mddefault}{\itdefault}k}}}}}
\put(1400,-500){\makebox(0,0)[lb]{\smash{{{\SetFigFont{8}{9.6}{\rmdefault}{\mddefault}{\rmdefault}(d)}}}}}
\end{picture}
}
\\
\end{array}\; =\; 
&
  \begin{array}[h]{c}
\setlength{\unitlength}{0.00033333in}
\begingroup\makeatletter\ifx\SetFigFont\undefined%
\gdef\SetFigFont#1#2#3#4#5{%
  \reset@font\fontsize{#1}{#2pt}%
  \fontfamily{#3}\fontseries{#4}\fontshape{#5}%
  \selectfont}%
\fi\endgroup%
{\renewcommand{\dashlinestretch}{30}
\begin{picture}(3328,2820)(0,-10)
\thicklines
\path(1698,357)(1698,493)
\path(1788,356)(1788,493)
\thinlines
\put(2258,1125){\makebox(0,0)[lb]{\smash{{{\SetFigFont{8}{9.6}{\rmdefault}{\mddefault}{\itdefault}R}}}}}
\put(807,1124){\makebox(0,0)[lb]{\smash{{{\SetFigFont{8}{9.6}{\rmdefault}{\mddefault}{\itdefault}A}}}}}
\path(1691,2290)(1634,2232)(1579,2290)
\path(1636,2526)(1637,2436)
\path(1637,2346)(1636,2256)
\path(1637,2166)(1636,2076)
\path(1637,2706)(1637,2616)
\path(1637,1986)(1637,1896)
\path(414,277)(494,257)(470,180)
\path(238,111)(315,158)
\path(393,203)(472,246)
\path(549,293)(628,336)
\path(81,22)(159,68)
\path(705,383)(783,427)
\path(2100,964)(2188,940)(2211,1027)
\path(1557,492)(1492,428)(1558,363)
\path(2799,183)(2779,261)(2855,280)
\path(3031,113)(2953,157)
\path(2875,202)(2797,249)
\path(2719,292)(2641,339)
\path(3188,22)(3109,67)
\path(2563,382)(2485,427)
\path(785,425)(2487,425)
\path(785,425)(1636,1900)
\path(1636,1901)(2487,426)
\path(1299,1190)(1276,1279)(1190,1254)
\put(1745,2649){\makebox(0,0)[lb]{\smash{{{\SetFigFont{8}{9.6}{\rmdefault}{\mddefault}{\itdefault}p}}}}}
\put(1880,2539){\makebox(0,0)[lb]{\smash{{{\SetFigFont{5}{6.0}{\rmdefault}{\mddefault}{\itdefault}3}}}}}
\put(0,292){\makebox(0,0)[lb]{\smash{{{\SetFigFont{8}{9.6}{\rmdefault}{\mddefault}{\itdefault}p}}}}}
\put(3097,292){\makebox(0,0)[lb]{\smash{{{\SetFigFont{8}{9.6}{\rmdefault}{\mddefault}{\itdefault}p}}}}}
\put(130,187){\makebox(0,0)[lb]{\smash{{{\SetFigFont{5}{6.0}{\rmdefault}{\mddefault}{\itdefault}2}}}}}
\put(3252,197){\makebox(0,0)[lb]{\smash{{{\SetFigFont{5}{6.0}{\rmdefault}{\mddefault}{\itdefault}1}}}}}
\put(1550,50){\makebox(0,0)[lb]{\smash{{{\SetFigFont{8}{9.6}{\rmdefault}{\mddefault}{\itdefault}k}}}}}
\put(1400,-100){\makebox(0,0)[lb]{\smash{{{\SetFigFont{8}{9.6}{\rmdefault}{\mddefault}{\rmdefault}{$~$}}}}}}
\end{picture}
}
\\ 
\end{array} .
\end{eqnarray}
In order to obtain this result, we make use of the relations 
(\ref{propag}) and note that the integrands associated with
the above diagrams can be combined pairwise as follows
\begin{equation}
I_{(a)}+I_{(b)}\sim G_{++}^{(0)}(k+p_2)\left[
G_{++}^{(0)}(k-p_1)-G_{+-}^{(0)}(k-p_1)\right]=
G_{++}^{(0)}(k+p_2)G_{R}(k-p_1)
\end{equation}
\begin{equation}
I_{(c)}+I_{(d)}\sim -G_{-+}^{(0)}(k+p_2)\left[
G_{++}^{(0)}(k-p_1)-G_{+-}^{(0)}(k-p_1)\right]=
-G_{-+}^{(0)}(k+p_2)G_{R}(k-p_1) .
\end{equation}
Finally, adding all contributions, we find that:
\begin{equation}
\begin{array}{ll}
I_{(a)}+I_{(b)}+I_{(c)}+I_{(d)} & \sim  \left[
G_{++}^{(0)}(k+p_2)-G_{-+}^{(0)}(k+p_2)\right]G_{R}(k-p_1) \\
{} & =  G_{A}(k+p_2)G_{R}(k-p_1) .
\end{array}
\end{equation}
The result given in equation (\ref{three1cut}) coincides with that
associated with the forward scattering amplitude shown in 
Fig. (3b), which was 
obtained after analytic continuation from the imaginary time formalism.

The sum  of the  diagrams with  distinct  single fixed cut  propagator
would,  then,  correspond  to different permutations   of the external
lines and it is clear that the sum of all  such diagrams would exactly
coincide with the  diagrams obtained  in  the earlier section in   the
imaginary time formalism.    Therefore, this set  of diagrams uniquely
corresponds  to  the  retarded three  point   function  that would  be
obtained from the imaginary time formalism 
\begin{equation}
\Gamma_{R}^{(3)}  =  \Gamma_{+++}  + \Gamma_{+-+}   +  \Gamma_{++-}  +
\Gamma_{+--}\label{b2} \hspace{.2in} .
\end{equation}
We parenthetically remark here that this is  unique to the extent that
the retarded self-energy is unique.  In fact, we  know that because of
the constraints among various causal amplitudes, we can also write 
\begin{equation}
\Sigma_{R} = - \Sigma_{-+} - \Sigma_{--} 
\end{equation}
Similarly, for the retarded three point amplitude, we can also write 
\begin{equation}
\Gamma_{R}^{(3)} =   - \Gamma_{-++} -  \Gamma_{--+} -   \Gamma_{-+-} -
\Gamma_{---} \hspace{.2in} .
\end{equation}
However, the  definition that   contains the  amplitude  with  all \lq\lq +''
vertices (namely, the physical   amplitude) is uniquely given  by  Eq.
(\ref{b2}). Furthermore, by  relating  them to the forward  scattering
amplitude, we, naturally, have a simple way of calculating them. 

Let us next look at the retarded four point  amplitude. Once again, it
is easy to see that the sum  of the following graphs with 4-cut propagators 
vanishes:
\vfill\eject
\begin{eqnarray}\label{four4cuts}
  \begin{array}[h]{c}
\setlength{\unitlength}{0.00041667in}
\begingroup\makeatletter\ifx\SetFigFont\undefined%
\gdef\SetFigFont#1#2#3#4#5{%
  \reset@font\fontsize{#1}{#2pt}%
  \fontfamily{#3}\fontseries{#4}\fontshape{#5}%
  \selectfont}%
\fi\endgroup%
{\renewcommand{\dashlinestretch}{30}
\begin{picture}(2486,2209)(0,-10)
\path(1151,468)(1084,418)(1143,345)
\thicklines
\path(1266,347)(1254,481)
\path(1356,357)(1339,492)
\thinlines
\path(1859,1010)(1907,941)(1982,998)
\thicklines
\path(1989,1131)(1855,1119)
\path(1979,1221)(1844,1204)
\thinlines
\path(1296,1734)(1360,1787)(1298,1857)
\thicklines
\path(1169,1867)(1181,1733)
\path(1079,1857)(1096,1722)
\thinlines
\thicklines
\path(459,1057)(593,1069)
\path(469,967)(604,984)
\path(611,1583)(476,1583)
\path(541,1517)(541,1653)
\path(606,611)(471,611)
\path(536,677)(536,541)
\path(1834,621)(1969,621)
\path(1904,687)(1904,551)
\path(1839,1573)(1974,1573)
\path(1909,1507)(1909,1643)
\thinlines
\put(1225,1102){\ellipse{1400}{1400}}
\path(2044,1829)(1963,1828)(1965,1908)
\path(2172,2035)(2109,1971)
\path(2046,1907)(1981,1844)
\path(1918,1780)(1854,1717)
\path(2300,2162)(2236,2098)
\path(1791,1652)(1727,1589)
\path(401,355)(482,356)(480,276)
\path(273,149)(336,213)
\path(399,277)(464,340)
\path(527,404)(591,467)
\path(145,22)(209,86)
\path(654,532)(718,595)
\path(406,1839)(487,1838)(485,1918)
\path(278,2045)(341,1981)
\path(404,1917)(469,1854)
\path(532,1790)(596,1727)
\path(150,2172)(214,2108)
\path(659,1662)(723,1599)
\path(2039,365)(1958,366)(1960,286)
\path(2167,159)(2104,223)
\path(2041,287)(1976,350)
\path(1913,414)(1849,477)
\path(2295,32)(2231,96)
\path(1786,542)(1722,605)
\path(589,1178)(547,1253)(466,1202)
\put(5,287){\makebox(0,0)[lb]{\smash{{{\SetFigFont{10}{12.0}{\rmdefault}{\mddefault}{\itdefault}p}}}}}
\put(2305,289){\makebox(0,0)[lb]{\smash{{{\SetFigFont{10}{12.0}{\rmdefault}{\mddefault}{\itdefault}p}}}}}
\put(2255,1869){\makebox(0,0)[lb]{\smash{{{\SetFigFont{10}{12.0}{\rmdefault}{\mddefault}{\itdefault}p}}}}}
\put(0,1869){\makebox(0,0)[lb]{\smash{{{\SetFigFont{10}{12.0}{\rmdefault}{\mddefault}{\itdefault}p}}}}}
\put(110,1764){\makebox(0,0)[lb]{\smash{{{\SetFigFont{5}{6.0}{\rmdefault}{\mddefault}{\itdefault}3}}}}}
\put(105,177){\makebox(0,0)[lb]{\smash{{{\SetFigFont{5}{6.0}{\rmdefault}{\mddefault}{\itdefault}2}}}}}
\put(2375,1774){\makebox(0,0)[lb]{\smash{{{\SetFigFont{5}{6.0}{\rmdefault}{\mddefault}{\itdefault}4}}}}}
\put(2410,179){\makebox(0,0)[lb]{\smash{{{\SetFigFont{5}{6.0}{\rmdefault}{\mddefault}{\itdefault}1}}}}}
\end{picture}
}
\end{array}\; +\; 
  \begin{array}[h]{c}
\setlength{\unitlength}{0.00041667in}
\begingroup\makeatletter\ifx\SetFigFont\undefined%
\gdef\SetFigFont#1#2#3#4#5{%
  \reset@font\fontsize{#1}{#2pt}%
  \fontfamily{#3}\fontseries{#4}\fontshape{#5}%
  \selectfont}%
\fi\endgroup%
{\renewcommand{\dashlinestretch}{30}
\begin{picture}(2486,2209)(0,-10)
\path(1151,468)(1084,418)(1143,345)
\thicklines
\path(1266,347)(1254,481)
\path(1356,357)(1339,492)
\thinlines
\path(1859,1010)(1907,941)(1982,998)
\thicklines
\path(1989,1131)(1855,1119)
\path(1979,1221)(1844,1204)
\thinlines
\path(1296,1734)(1360,1787)(1298,1857)
\thicklines
\path(1169,1867)(1181,1733)
\path(1079,1857)(1096,1722)
\thinlines
\thicklines
\path(459,1057)(593,1069)
\path(469,967)(604,984)
\path(611,1583)(476,1583)
\path(606,611)(471,611)
\path(536,677)(536,541)
\path(1834,621)(1969,621)
\path(1904,687)(1904,551)
\path(1839,1573)(1974,1573)
\path(1909,1507)(1909,1643)
\thinlines
\put(1225,1102){\ellipse{1400}{1400}}
\path(2044,1829)(1963,1828)(1965,1908)
\path(2172,2035)(2109,1971)
\path(2046,1907)(1981,1844)
\path(1918,1780)(1854,1717)
\path(2300,2162)(2236,2098)
\path(1791,1652)(1727,1589)
\path(401,355)(482,356)(480,276)
\path(273,149)(336,213)
\path(399,277)(464,340)
\path(527,404)(591,467)
\path(145,22)(209,86)
\path(654,532)(718,595)
\path(406,1839)(487,1838)(485,1918)
\path(278,2045)(341,1981)
\path(404,1917)(469,1854)
\path(532,1790)(596,1727)
\path(150,2172)(214,2108)
\path(659,1662)(723,1599)
\path(2039,365)(1958,366)(1960,286)
\path(2167,159)(2104,223)
\path(2041,287)(1976,350)
\path(1913,414)(1849,477)
\path(2295,32)(2231,96)
\path(1786,542)(1722,605)
\path(589,1178)(547,1253)(466,1202)
\put(5,287){\makebox(0,0)[lb]{\smash{{{\SetFigFont{10}{12.0}{\rmdefault}{\mddefault}{\itdefault}p}}}}}
\put(2305,289){\makebox(0,0)[lb]{\smash{{{\SetFigFont{10}{12.0}{\rmdefault}{\mddefault}{\itdefault}p}}}}}
\put(2255,1869){\makebox(0,0)[lb]{\smash{{{\SetFigFont{10}{12.0}{\rmdefault}{\mddefault}{\itdefault}p}}}}}
\put(0,1869){\makebox(0,0)[lb]{\smash{{{\SetFigFont{10}{12.0}{\rmdefault}{\mddefault}{\itdefault}p}}}}}
\put(110,1764){\makebox(0,0)[lb]{\smash{{{\SetFigFont{5}{6.0}{\rmdefault}{\mddefault}{\itdefault}3}}}}}
\put(105,177){\makebox(0,0)[lb]{\smash{{{\SetFigFont{5}{6.0}{\rmdefault}{\mddefault}{\itdefault}2}}}}}
\put(2375,1774){\makebox(0,0)[lb]{\smash{{{\SetFigFont{5}{6.0}{\rmdefault}{\mddefault}{\itdefault}4}}}}}
\put(2410,179){\makebox(0,0)[lb]{\smash{{{\SetFigFont{5}{6.0}{\rmdefault}{\mddefault}{\itdefault}1}}}}}
\end{picture}
}
\end{array}\; +\;   
  \begin{array}[h]{c}
\setlength{\unitlength}{0.00041667in}
\begingroup\makeatletter\ifx\SetFigFont\undefined%
\gdef\SetFigFont#1#2#3#4#5{%
  \reset@font\fontsize{#1}{#2pt}%
  \fontfamily{#3}\fontseries{#4}\fontshape{#5}%
  \selectfont}%
\fi\endgroup%
{\renewcommand{\dashlinestretch}{30}
\begin{picture}(2486,2209)(0,-10)
\path(1151,468)(1084,418)(1143,345)
\thicklines
\path(1266,347)(1254,481)
\path(1356,357)(1339,492)
\thinlines
\path(1859,1010)(1907,941)(1982,998)
\thicklines
\path(1989,1131)(1855,1119)
\path(1979,1221)(1844,1204)
\thinlines
\path(1296,1734)(1360,1787)(1298,1857)
\thicklines
\path(1169,1867)(1181,1733)
\path(1079,1857)(1096,1722)
\thinlines
\thicklines
\path(459,1057)(593,1069)
\path(469,967)(604,984)
\path(611,1583)(476,1583)
\path(541,1517)(541,1653)
\path(606,611)(471,611)
\path(536,677)(536,541)
\path(1834,621)(1969,621)
\path(1839,1573)(1974,1573)
\path(1909,1507)(1909,1643)
\thinlines
\put(1225,1102){\ellipse{1400}{1400}}
\path(2044,1829)(1963,1828)(1965,1908)
\path(2172,2035)(2109,1971)
\path(2046,1907)(1981,1844)
\path(1918,1780)(1854,1717)
\path(2300,2162)(2236,2098)
\path(1791,1652)(1727,1589)
\path(401,355)(482,356)(480,276)
\path(273,149)(336,213)
\path(399,277)(464,340)
\path(527,404)(591,467)
\path(145,22)(209,86)
\path(654,532)(718,595)
\path(406,1839)(487,1838)(485,1918)
\path(278,2045)(341,1981)
\path(404,1917)(469,1854)
\path(532,1790)(596,1727)
\path(150,2172)(214,2108)
\path(659,1662)(723,1599)
\path(2039,365)(1958,366)(1960,286)
\path(2167,159)(2104,223)
\path(2041,287)(1976,350)
\path(1913,414)(1849,477)
\path(2295,32)(2231,96)
\path(1786,542)(1722,605)
\path(589,1178)(547,1253)(466,1202)
\put(5,287){\makebox(0,0)[lb]{\smash{{{\SetFigFont{10}{12.0}{\rmdefault}{\mddefault}{\itdefault}p}}}}}
\put(2305,289){\makebox(0,0)[lb]{\smash{{{\SetFigFont{10}{12.0}{\rmdefault}{\mddefault}{\itdefault}p}}}}}
\put(2255,1869){\makebox(0,0)[lb]{\smash{{{\SetFigFont{10}{12.0}{\rmdefault}{\mddefault}{\itdefault}p}}}}}
\put(0,1869){\makebox(0,0)[lb]{\smash{{{\SetFigFont{10}{12.0}{\rmdefault}{\mddefault}{\itdefault}p}}}}}
\put(110,1764){\makebox(0,0)[lb]{\smash{{{\SetFigFont{5}{6.0}{\rmdefault}{\mddefault}{\itdefault}3}}}}}
\put(105,177){\makebox(0,0)[lb]{\smash{{{\SetFigFont{5}{6.0}{\rmdefault}{\mddefault}{\itdefault}2}}}}}
\put(2375,1774){\makebox(0,0)[lb]{\smash{{{\SetFigFont{5}{6.0}{\rmdefault}{\mddefault}{\itdefault}4}}}}}
\put(2410,179){\makebox(0,0)[lb]{\smash{{{\SetFigFont{5}{6.0}{\rmdefault}{\mddefault}{\itdefault}1}}}}}
\end{picture}
}
\end{array}\; +\; 
  \begin{array}[h]{c}
\setlength{\unitlength}{0.00041667in}
\begingroup\makeatletter\ifx\SetFigFont\undefined%
\gdef\SetFigFont#1#2#3#4#5{%
  \reset@font\fontsize{#1}{#2pt}%
  \fontfamily{#3}\fontseries{#4}\fontshape{#5}%
  \selectfont}%
\fi\endgroup%
{\renewcommand{\dashlinestretch}{30}
\begin{picture}(2486,2209)(0,-10)
\path(1151,468)(1084,418)(1143,345)
\thicklines
\path(1266,347)(1254,481)
\path(1356,357)(1339,492)
\thinlines
\path(1859,1010)(1907,941)(1982,998)
\thicklines
\path(1989,1131)(1855,1119)
\path(1979,1221)(1844,1204)
\thinlines
\path(1296,1734)(1360,1787)(1298,1857)
\thicklines
\path(1169,1867)(1181,1733)
\path(1079,1857)(1096,1722)
\thinlines
\thicklines
\path(459,1057)(593,1069)
\path(469,967)(604,984)
\path(611,1583)(476,1583)
\path(606,611)(471,611)
\path(536,677)(536,541)
\path(1834,621)(1969,621)
\path(1839,1573)(1974,1573)
\path(1909,1507)(1909,1643)
\thinlines
\put(1225,1102){\ellipse{1400}{1400}}
\path(2044,1829)(1963,1828)(1965,1908)
\path(2172,2035)(2109,1971)
\path(2046,1907)(1981,1844)
\path(1918,1780)(1854,1717)
\path(2300,2162)(2236,2098)
\path(1791,1652)(1727,1589)
\path(401,355)(482,356)(480,276)
\path(273,149)(336,213)
\path(399,277)(464,340)
\path(527,404)(591,467)
\path(145,22)(209,86)
\path(654,532)(718,595)
\path(406,1839)(487,1838)(485,1918)
\path(278,2045)(341,1981)
\path(404,1917)(469,1854)
\path(532,1790)(596,1727)
\path(150,2172)(214,2108)
\path(659,1662)(723,1599)
\path(2039,365)(1958,366)(1960,286)
\path(2167,159)(2104,223)
\path(2041,287)(1976,350)
\path(1913,414)(1849,477)
\path(2295,32)(2231,96)
\path(1786,542)(1722,605)
\path(589,1178)(547,1253)(466,1202)
\put(5,287){\makebox(0,0)[lb]{\smash{{{\SetFigFont{10}{12.0}{\rmdefault}{\mddefault}{\itdefault}p}}}}}
\put(2305,289){\makebox(0,0)[lb]{\smash{{{\SetFigFont{10}{12.0}{\rmdefault}{\mddefault}{\itdefault}p}}}}}
\put(2255,1869){\makebox(0,0)[lb]{\smash{{{\SetFigFont{10}{12.0}{\rmdefault}{\mddefault}{\itdefault}p}}}}}
\put(0,1869){\makebox(0,0)[lb]{\smash{{{\SetFigFont{10}{12.0}{\rmdefault}{\mddefault}{\itdefault}p}}}}}
\put(110,1764){\makebox(0,0)[lb]{\smash{{{\SetFigFont{5}{6.0}{\rmdefault}{\mddefault}{\itdefault}3}}}}}
\put(105,177){\makebox(0,0)[lb]{\smash{{{\SetFigFont{5}{6.0}{\rmdefault}{\mddefault}{\itdefault}2}}}}}
\put(2375,1774){\makebox(0,0)[lb]{\smash{{{\SetFigFont{5}{6.0}{\rmdefault}{\mddefault}{\itdefault}4}}}}}
\put(2410,179){\makebox(0,0)[lb]{\smash{{{\SetFigFont{5}{6.0}{\rmdefault}{\mddefault}{\itdefault}1}}}}}
\end{picture}
}
\end{array}\; +\; 
& {} & {} \nonumber \\
& {} & {} \nonumber \\
  \begin{array}[h]{c} 
\setlength{\unitlength}{0.00041667in}
\begingroup\makeatletter\ifx\SetFigFont\undefined%
\gdef\SetFigFont#1#2#3#4#5{%
  \reset@font\fontsize{#1}{#2pt}%
  \fontfamily{#3}\fontseries{#4}\fontshape{#5}%
  \selectfont}%
\fi\endgroup%
{\renewcommand{\dashlinestretch}{30}
\begin{picture}(2486,2209)(0,-10)
\path(1151,468)(1084,418)(1143,345)
\thicklines
\path(1266,347)(1254,481)
\path(1356,357)(1339,492)
\thinlines
\path(1859,1010)(1907,941)(1982,998)
\thicklines
\path(1989,1131)(1855,1119)
\path(1979,1221)(1844,1204)
\thinlines
\path(1296,1734)(1360,1787)(1298,1857)
\thicklines
\path(1169,1867)(1181,1733)
\path(1079,1857)(1096,1722)
\thinlines
\thicklines
\path(459,1057)(593,1069)
\path(469,967)(604,984)
\path(611,1583)(476,1583)
\path(541,1517)(541,1653)
\path(606,611)(471,611)
\path(1834,621)(1969,621)
\path(1904,687)(1904,551)
\path(1839,1573)(1974,1573)
\path(1909,1507)(1909,1643)
\thinlines
\put(1225,1102){\ellipse{1400}{1400}}
\path(2044,1829)(1963,1828)(1965,1908)
\path(2172,2035)(2109,1971)
\path(2046,1907)(1981,1844)
\path(1918,1780)(1854,1717)
\path(2300,2162)(2236,2098)
\path(1791,1652)(1727,1589)
\path(401,355)(482,356)(480,276)
\path(273,149)(336,213)
\path(399,277)(464,340)
\path(527,404)(591,467)
\path(145,22)(209,86)
\path(654,532)(718,595)
\path(406,1839)(487,1838)(485,1918)
\path(278,2045)(341,1981)
\path(404,1917)(469,1854)
\path(532,1790)(596,1727)
\path(150,2172)(214,2108)
\path(659,1662)(723,1599)
\path(2039,365)(1958,366)(1960,286)
\path(2167,159)(2104,223)
\path(2041,287)(1976,350)
\path(1913,414)(1849,477)
\path(2295,32)(2231,96)
\path(1786,542)(1722,605)
\path(589,1178)(547,1253)(466,1202)
\put(5,287){\makebox(0,0)[lb]{\smash{{{\SetFigFont{10}{12.0}{\rmdefault}{\mddefault}{\itdefault}p}}}}}
\put(2305,289){\makebox(0,0)[lb]{\smash{{{\SetFigFont{10}{12.0}{\rmdefault}{\mddefault}{\itdefault}p}}}}}
\put(2255,1869){\makebox(0,0)[lb]{\smash{{{\SetFigFont{10}{12.0}{\rmdefault}{\mddefault}{\itdefault}p}}}}}
\put(0,1869){\makebox(0,0)[lb]{\smash{{{\SetFigFont{10}{12.0}{\rmdefault}{\mddefault}{\itdefault}p}}}}}
\put(110,1764){\makebox(0,0)[lb]{\smash{{{\SetFigFont{5}{6.0}{\rmdefault}{\mddefault}{\itdefault}3}}}}}
\put(105,177){\makebox(0,0)[lb]{\smash{{{\SetFigFont{5}{6.0}{\rmdefault}{\mddefault}{\itdefault}2}}}}}
\put(2375,1774){\makebox(0,0)[lb]{\smash{{{\SetFigFont{5}{6.0}{\rmdefault}{\mddefault}{\itdefault}4}}}}}
\put(2410,179){\makebox(0,0)[lb]{\smash{{{\SetFigFont{5}{6.0}{\rmdefault}{\mddefault}{\itdefault}1}}}}}
\end{picture}
}
\end{array}\; +\; 
  \begin{array}[h]{c}
\setlength{\unitlength}{0.00041667in}
\begingroup\makeatletter\ifx\SetFigFont\undefined%
\gdef\SetFigFont#1#2#3#4#5{%
  \reset@font\fontsize{#1}{#2pt}%
  \fontfamily{#3}\fontseries{#4}\fontshape{#5}%
  \selectfont}%
\fi\endgroup%
{\renewcommand{\dashlinestretch}{30}
\begin{picture}(2486,2209)(0,-10)
\path(1151,468)(1084,418)(1143,345)
\thicklines
\path(1266,347)(1254,481)
\path(1356,357)(1339,492)
\thinlines
\path(1859,1010)(1907,941)(1982,998)
\thicklines
\path(1989,1131)(1855,1119)
\path(1979,1221)(1844,1204)
\thinlines
\path(1296,1734)(1360,1787)(1298,1857)
\thicklines
\path(1169,1867)(1181,1733)
\path(1079,1857)(1096,1722)
\thinlines
\thicklines
\path(459,1057)(593,1069)
\path(469,967)(604,984)
\path(611,1583)(476,1583)
\path(606,611)(471,611)
\path(1834,621)(1969,621)
\path(1904,687)(1904,551)
\path(1839,1573)(1974,1573)
\path(1909,1507)(1909,1643)
\thinlines
\put(1225,1102){\ellipse{1400}{1400}}
\path(2044,1829)(1963,1828)(1965,1908)
\path(2172,2035)(2109,1971)
\path(2046,1907)(1981,1844)
\path(1918,1780)(1854,1717)
\path(2300,2162)(2236,2098)
\path(1791,1652)(1727,1589)
\path(401,355)(482,356)(480,276)
\path(273,149)(336,213)
\path(399,277)(464,340)
\path(527,404)(591,467)
\path(145,22)(209,86)
\path(654,532)(718,595)
\path(406,1839)(487,1838)(485,1918)
\path(278,2045)(341,1981)
\path(404,1917)(469,1854)
\path(532,1790)(596,1727)
\path(150,2172)(214,2108)
\path(659,1662)(723,1599)
\path(2039,365)(1958,366)(1960,286)
\path(2167,159)(2104,223)
\path(2041,287)(1976,350)
\path(1913,414)(1849,477)
\path(2295,32)(2231,96)
\path(1786,542)(1722,605)
\path(589,1178)(547,1253)(466,1202)
\put(5,287){\makebox(0,0)[lb]{\smash{{{\SetFigFont{10}{12.0}{\rmdefault}{\mddefault}{\itdefault}p}}}}}
\put(2305,289){\makebox(0,0)[lb]{\smash{{{\SetFigFont{10}{12.0}{\rmdefault}{\mddefault}{\itdefault}p}}}}}
\put(2255,1869){\makebox(0,0)[lb]{\smash{{{\SetFigFont{10}{12.0}{\rmdefault}{\mddefault}{\itdefault}p}}}}}
\put(0,1869){\makebox(0,0)[lb]{\smash{{{\SetFigFont{10}{12.0}{\rmdefault}{\mddefault}{\itdefault}p}}}}}
\put(110,1764){\makebox(0,0)[lb]{\smash{{{\SetFigFont{5}{6.0}{\rmdefault}{\mddefault}{\itdefault}3}}}}}
\put(105,177){\makebox(0,0)[lb]{\smash{{{\SetFigFont{5}{6.0}{\rmdefault}{\mddefault}{\itdefault}2}}}}}
\put(2375,1774){\makebox(0,0)[lb]{\smash{{{\SetFigFont{5}{6.0}{\rmdefault}{\mddefault}{\itdefault}4}}}}}
\put(2410,179){\makebox(0,0)[lb]{\smash{{{\SetFigFont{5}{6.0}{\rmdefault}{\mddefault}{\itdefault}1}}}}}
\end{picture}
}
\end{array}\; +\;
  \begin{array}[h]{c}
\setlength{\unitlength}{0.00041667in}
\begingroup\makeatletter\ifx\SetFigFont\undefined%
\gdef\SetFigFont#1#2#3#4#5{%
  \reset@font\fontsize{#1}{#2pt}%
  \fontfamily{#3}\fontseries{#4}\fontshape{#5}%
  \selectfont}%
\fi\endgroup%
{\renewcommand{\dashlinestretch}{30}
\begin{picture}(2486,2209)(0,-10)
\path(1151,468)(1084,418)(1143,345)
\thicklines
\path(1266,347)(1254,481)
\path(1356,357)(1339,492)
\thinlines
\path(1859,1010)(1907,941)(1982,998)
\thicklines
\path(1989,1131)(1855,1119)
\path(1979,1221)(1844,1204)
\thinlines
\path(1296,1734)(1360,1787)(1298,1857)
\thicklines
\path(1169,1867)(1181,1733)
\path(1079,1857)(1096,1722)
\thinlines
\thicklines
\path(459,1057)(593,1069)
\path(469,967)(604,984)
\path(611,1583)(476,1583)
\path(541,1517)(541,1653)
\path(606,611)(471,611)
\path(1834,621)(1969,621)
\path(1839,1573)(1974,1573)
\path(1909,1507)(1909,1643)
\thinlines
\put(1225,1102){\ellipse{1400}{1400}}
\path(2044,1829)(1963,1828)(1965,1908)
\path(2172,2035)(2109,1971)
\path(2046,1907)(1981,1844)
\path(1918,1780)(1854,1717)
\path(2300,2162)(2236,2098)
\path(1791,1652)(1727,1589)
\path(401,355)(482,356)(480,276)
\path(273,149)(336,213)
\path(399,277)(464,340)
\path(527,404)(591,467)
\path(145,22)(209,86)
\path(654,532)(718,595)
\path(406,1839)(487,1838)(485,1918)
\path(278,2045)(341,1981)
\path(404,1917)(469,1854)
\path(532,1790)(596,1727)
\path(150,2172)(214,2108)
\path(659,1662)(723,1599)
\path(2039,365)(1958,366)(1960,286)
\path(2167,159)(2104,223)
\path(2041,287)(1976,350)
\path(1913,414)(1849,477)
\path(2295,32)(2231,96)
\path(1786,542)(1722,605)
\path(589,1178)(547,1253)(466,1202)
\put(5,287){\makebox(0,0)[lb]{\smash{{{\SetFigFont{10}{12.0}{\rmdefault}{\mddefault}{\itdefault}p}}}}}
\put(2305,289){\makebox(0,0)[lb]{\smash{{{\SetFigFont{10}{12.0}{\rmdefault}{\mddefault}{\itdefault}p}}}}}
\put(2255,1869){\makebox(0,0)[lb]{\smash{{{\SetFigFont{10}{12.0}{\rmdefault}{\mddefault}{\itdefault}p}}}}}
\put(0,1869){\makebox(0,0)[lb]{\smash{{{\SetFigFont{10}{12.0}{\rmdefault}{\mddefault}{\itdefault}p}}}}}
\put(110,1764){\makebox(0,0)[lb]{\smash{{{\SetFigFont{5}{6.0}{\rmdefault}{\mddefault}{\itdefault}3}}}}}
\put(105,177){\makebox(0,0)[lb]{\smash{{{\SetFigFont{5}{6.0}{\rmdefault}{\mddefault}{\itdefault}2}}}}}
\put(2375,1774){\makebox(0,0)[lb]{\smash{{{\SetFigFont{5}{6.0}{\rmdefault}{\mddefault}{\itdefault}4}}}}}
\put(2410,179){\makebox(0,0)[lb]{\smash{{{\SetFigFont{5}{6.0}{\rmdefault}{\mddefault}{\itdefault}1}}}}}
\end{picture}
}
\end{array}\; +\; 
  \begin{array}[h]{c}
\setlength{\unitlength}{0.00041667in}
\begingroup\makeatletter\ifx\SetFigFont\undefined%
\gdef\SetFigFont#1#2#3#4#5{%
  \reset@font\fontsize{#1}{#2pt}%
  \fontfamily{#3}\fontseries{#4}\fontshape{#5}%
  \selectfont}%
\fi\endgroup%
{\renewcommand{\dashlinestretch}{30}
\begin{picture}(2486,2209)(0,-10)
\path(1151,468)(1084,418)(1143,345)
\thicklines
\path(1266,347)(1254,481)
\path(1356,357)(1339,492)
\thinlines
\path(1859,1010)(1907,941)(1982,998)
\thicklines
\path(1989,1131)(1855,1119)
\path(1979,1221)(1844,1204)
\thinlines
\path(1296,1734)(1360,1787)(1298,1857)
\thicklines
\path(1169,1867)(1181,1733)
\path(1079,1857)(1096,1722)
\thinlines
\thicklines
\path(459,1057)(593,1069)
\path(469,967)(604,984)
\path(611,1583)(476,1583)
\path(606,611)(471,611)
\path(1834,621)(1969,621)
\path(1839,1573)(1974,1573)
\path(1909,1507)(1909,1643)
\thinlines
\put(1225,1102){\ellipse{1400}{1400}}
\path(2044,1829)(1963,1828)(1965,1908)
\path(2172,2035)(2109,1971)
\path(2046,1907)(1981,1844)
\path(1918,1780)(1854,1717)
\path(2300,2162)(2236,2098)
\path(1791,1652)(1727,1589)
\path(401,355)(482,356)(480,276)
\path(273,149)(336,213)
\path(399,277)(464,340)
\path(527,404)(591,467)
\path(145,22)(209,86)
\path(654,532)(718,595)
\path(406,1839)(487,1838)(485,1918)
\path(278,2045)(341,1981)
\path(404,1917)(469,1854)
\path(532,1790)(596,1727)
\path(150,2172)(214,2108)
\path(659,1662)(723,1599)
\path(2039,365)(1958,366)(1960,286)
\path(2167,159)(2104,223)
\path(2041,287)(1976,350)
\path(1913,414)(1849,477)
\path(2295,32)(2231,96)
\path(1786,542)(1722,605)
\path(589,1178)(547,1253)(466,1202)
\put(5,287){\makebox(0,0)[lb]{\smash{{{\SetFigFont{10}{12.0}{\rmdefault}{\mddefault}{\itdefault}p}}}}}
\put(2305,289){\makebox(0,0)[lb]{\smash{{{\SetFigFont{10}{12.0}{\rmdefault}{\mddefault}{\itdefault}p}}}}}
\put(2255,1869){\makebox(0,0)[lb]{\smash{{{\SetFigFont{10}{12.0}{\rmdefault}{\mddefault}{\itdefault}p}}}}}
\put(0,1869){\makebox(0,0)[lb]{\smash{{{\SetFigFont{10}{12.0}{\rmdefault}{\mddefault}{\itdefault}p}}}}}
\put(110,1764){\makebox(0,0)[lb]{\smash{{{\SetFigFont{5}{6.0}{\rmdefault}{\mddefault}{\itdefault}3}}}}}
\put(105,177){\makebox(0,0)[lb]{\smash{{{\SetFigFont{5}{6.0}{\rmdefault}{\mddefault}{\itdefault}2}}}}}
\put(2375,1774){\makebox(0,0)[lb]{\smash{{{\SetFigFont{5}{6.0}{\rmdefault}{\mddefault}{\itdefault}4}}}}}
\put(2410,179){\makebox(0,0)[lb]{\smash{{{\SetFigFont{5}{6.0}{\rmdefault}{\mddefault}{\itdefault}1}}}}}
\end{picture}
}
\end{array}\; {~} \;  
& {} &
=  \begin{array}[h]{c} 0 \end{array} .
\end{eqnarray}
Similarly, it can be verified that the sum  of 3-cut propagators as well as
the sum  of 2-cut propagators also vanish.
However, the diagrams  with a single cut  propagator do not add up  to
zero. Thus, it is clear that this set of diagrams is likely to lead to
the  retarded  four point  amplitude that  will  be obtained  from the
imaginary time formalism. Once  again, it is  easy to see that the sum
of the  diagrams with a fixed  single  cut propagator  correspond to a
given forward scattering amplitude and  that distinct fixed single cut
propagators  would correspond  to cyclic permutations  of the external
lines. For example, with the help of the relations (\ref{propag}),
one finds that
\begin{eqnarray}\label{four1cut}
\begin{array}[h]{c}
\setlength{\unitlength}{0.00041667in}
\begingroup\makeatletter\ifx\SetFigFont\undefined%
\gdef\SetFigFont#1#2#3#4#5{%
  \reset@font\fontsize{#1}{#2pt}%
  \fontfamily{#3}\fontseries{#4}\fontshape{#5}%
  \selectfont}%
\fi\endgroup%
{\renewcommand{\dashlinestretch}{30}
\begin{picture}(2486,2209)(0,-10)
\thicklines
\path(459,1057)(593,1069)
\path(469,967)(604,984)
\path(611,1583)(476,1583)
\path(541,1517)(541,1653)
\path(606,611)(471,611)
\path(536,677)(536,541)
\path(1834,621)(1969,621)
\path(1904,687)(1904,551)
\path(1839,1573)(1974,1573)
\path(1909,1507)(1909,1643)
\thinlines
\put(1225,1102){\ellipse{1400}{1400}}
\path(2044,1829)(1963,1828)(1965,1908)
\path(2172,2035)(2109,1971)
\path(2046,1907)(1981,1844)
\path(1918,1780)(1854,1717)
\path(2300,2162)(2236,2098)
\path(1791,1652)(1727,1589)
\path(401,355)(482,356)(480,276)
\path(273,149)(336,213)
\path(399,277)(464,340)
\path(527,404)(591,467)
\path(145,22)(209,86)
\path(654,532)(718,595)
\path(406,1839)(487,1838)(485,1918)
\path(278,2045)(341,1981)
\path(404,1917)(469,1854)
\path(532,1790)(596,1727)
\path(150,2172)(214,2108)
\path(659,1662)(723,1599)
\path(2039,365)(1958,366)(1960,286)
\path(2167,159)(2104,223)
\path(2041,287)(1976,350)
\path(1913,414)(1849,477)
\path(2295,32)(2231,96)
\path(1786,542)(1722,605)
\path(589,1178)(547,1253)(466,1202)
\put(5,287){\makebox(0,0)[lb]{\smash{{{\SetFigFont{10}{12.0}{\rmdefault}{\mddefault}{\itdefault}p}}}}}
\put(2305,289){\makebox(0,0)[lb]{\smash{{{\SetFigFont{10}{12.0}{\rmdefault}{\mddefault}{\itdefault}p}}}}}
\put(2255,1869){\makebox(0,0)[lb]{\smash{{{\SetFigFont{10}{12.0}{\rmdefault}{\mddefault}{\itdefault}p}}}}}
\put(0,1869){\makebox(0,0)[lb]{\smash{{{\SetFigFont{10}{12.0}{\rmdefault}{\mddefault}{\itdefault}p}}}}}
\put(110,1764){\makebox(0,0)[lb]{\smash{{{\SetFigFont{5}{6.0}{\rmdefault}{\mddefault}{\itdefault}3}}}}}
\put(105,177){\makebox(0,0)[lb]{\smash{{{\SetFigFont{5}{6.0}{\rmdefault}{\mddefault}{\itdefault}2}}}}}
\put(2375,1774){\makebox(0,0)[lb]{\smash{{{\SetFigFont{5}{6.0}{\rmdefault}{\mddefault}{\itdefault}4}}}}}
\put(2410,179){\makebox(0,0)[lb]{\smash{{{\SetFigFont{5}{6.0}{\rmdefault}{\mddefault}{\itdefault}1}}}}}
\end{picture}
}
  \end{array}\; +\; 
&
  \begin{array}[h]{c}
\setlength{\unitlength}{0.00041667in}
\begingroup\makeatletter\ifx\SetFigFont\undefined%
\gdef\SetFigFont#1#2#3#4#5{%
  \reset@font\fontsize{#1}{#2pt}%
  \fontfamily{#3}\fontseries{#4}\fontshape{#5}%
  \selectfont}%
\fi\endgroup%
{\renewcommand{\dashlinestretch}{30}
\begin{picture}(2486,2209)(0,-10)
\thicklines
\path(459,1057)(593,1069)
\path(469,967)(604,984)
\path(611,1583)(476,1583)
\path(606,611)(471,611)
\path(536,677)(536,541)
\path(1834,621)(1969,621)
\path(1904,687)(1904,551)
\path(1839,1573)(1974,1573)
\path(1909,1507)(1909,1643)
\thinlines
\put(1225,1102){\ellipse{1400}{1400}}
\path(2044,1829)(1963,1828)(1965,1908)
\path(2172,2035)(2109,1971)
\path(2046,1907)(1981,1844)
\path(1918,1780)(1854,1717)
\path(2300,2162)(2236,2098)
\path(1791,1652)(1727,1589)
\path(401,355)(482,356)(480,276)
\path(273,149)(336,213)
\path(399,277)(464,340)
\path(527,404)(591,467)
\path(145,22)(209,86)
\path(654,532)(718,595)
\path(406,1839)(487,1838)(485,1918)
\path(278,2045)(341,1981)
\path(404,1917)(469,1854)
\path(532,1790)(596,1727)
\path(150,2172)(214,2108)
\path(659,1662)(723,1599)
\path(2039,365)(1958,366)(1960,286)
\path(2167,159)(2104,223)
\path(2041,287)(1976,350)
\path(1913,414)(1849,477)
\path(2295,32)(2231,96)
\path(1786,542)(1722,605)
\path(589,1178)(547,1253)(466,1202)
\put(5,287){\makebox(0,0)[lb]{\smash{{{\SetFigFont{10}{12.0}{\rmdefault}{\mddefault}{\itdefault}p}}}}}
\put(2305,289){\makebox(0,0)[lb]{\smash{{{\SetFigFont{10}{12.0}{\rmdefault}{\mddefault}{\itdefault}p}}}}}
\put(2255,1869){\makebox(0,0)[lb]{\smash{{{\SetFigFont{10}{12.0}{\rmdefault}{\mddefault}{\itdefault}p}}}}}
\put(0,1869){\makebox(0,0)[lb]{\smash{{{\SetFigFont{10}{12.0}{\rmdefault}{\mddefault}{\itdefault}p}}}}}
\put(110,1764){\makebox(0,0)[lb]{\smash{{{\SetFigFont{5}{6.0}{\rmdefault}{\mddefault}{\itdefault}3}}}}}
\put(105,177){\makebox(0,0)[lb]{\smash{{{\SetFigFont{5}{6.0}{\rmdefault}{\mddefault}{\itdefault}2}}}}}
\put(2375,1774){\makebox(0,0)[lb]{\smash{{{\SetFigFont{5}{6.0}{\rmdefault}{\mddefault}{\itdefault}4}}}}}
\put(2410,179){\makebox(0,0)[lb]{\smash{{{\SetFigFont{5}{6.0}{\rmdefault}{\mddefault}{\itdefault}1}}}}}
\end{picture}
}
  \end{array}
\; +\;  
  \begin{array}[h]{c}
\setlength{\unitlength}{0.00041667in}
\begingroup\makeatletter\ifx\SetFigFont\undefined%
\gdef\SetFigFont#1#2#3#4#5{%
  \reset@font\fontsize{#1}{#2pt}%
  \fontfamily{#3}\fontseries{#4}\fontshape{#5}%
  \selectfont}%
\fi\endgroup%
{\renewcommand{\dashlinestretch}{30}
\begin{picture}(2486,2209)(0,-10)
\thicklines
\path(459,1057)(593,1069)
\path(469,967)(604,984)
\path(611,1583)(476,1583)
\path(541,1517)(541,1653)
\path(606,611)(471,611)
\path(536,677)(536,541)
\path(1834,621)(1969,621)
\path(1839,1573)(1974,1573)
\path(1909,1507)(1909,1643)
\thinlines
\put(1225,1102){\ellipse{1400}{1400}}
\path(2044,1829)(1963,1828)(1965,1908)
\path(2172,2035)(2109,1971)
\path(2046,1907)(1981,1844)
\path(1918,1780)(1854,1717)
\path(2300,2162)(2236,2098)
\path(1791,1652)(1727,1589)
\path(401,355)(482,356)(480,276)
\path(273,149)(336,213)
\path(399,277)(464,340)
\path(527,404)(591,467)
\path(145,22)(209,86)
\path(654,532)(718,595)
\path(406,1839)(487,1838)(485,1918)
\path(278,2045)(341,1981)
\path(404,1917)(469,1854)
\path(532,1790)(596,1727)
\path(150,2172)(214,2108)
\path(659,1662)(723,1599)
\path(2039,365)(1958,366)(1960,286)
\path(2167,159)(2104,223)
\path(2041,287)(1976,350)
\path(1913,414)(1849,477)
\path(2295,32)(2231,96)
\path(1786,542)(1722,605)
\path(589,1178)(547,1253)(466,1202)
\put(5,287){\makebox(0,0)[lb]{\smash{{{\SetFigFont{10}{12.0}{\rmdefault}{\mddefault}{\itdefault}p}}}}}
\put(2305,289){\makebox(0,0)[lb]{\smash{{{\SetFigFont{10}{12.0}{\rmdefault}{\mddefault}{\itdefault}p}}}}}
\put(2255,1869){\makebox(0,0)[lb]{\smash{{{\SetFigFont{10}{12.0}{\rmdefault}{\mddefault}{\itdefault}p}}}}}
\put(0,1869){\makebox(0,0)[lb]{\smash{{{\SetFigFont{10}{12.0}{\rmdefault}{\mddefault}{\itdefault}p}}}}}
\put(110,1764){\makebox(0,0)[lb]{\smash{{{\SetFigFont{5}{6.0}{\rmdefault}{\mddefault}{\itdefault}3}}}}}
\put(105,177){\makebox(0,0)[lb]{\smash{{{\SetFigFont{5}{6.0}{\rmdefault}{\mddefault}{\itdefault}2}}}}}
\put(2375,1774){\makebox(0,0)[lb]{\smash{{{\SetFigFont{5}{6.0}{\rmdefault}{\mddefault}{\itdefault}4}}}}}
\put(2410,179){\makebox(0,0)[lb]{\smash{{{\SetFigFont{5}{6.0}{\rmdefault}{\mddefault}{\itdefault}1}}}}}
\end{picture}
}
  \end{array}
\; +\; 
&
  \begin{array}[h]{c}
\setlength{\unitlength}{0.00041667in}
\begingroup\makeatletter\ifx\SetFigFont\undefined%
\gdef\SetFigFont#1#2#3#4#5{%
  \reset@font\fontsize{#1}{#2pt}%
  \fontfamily{#3}\fontseries{#4}\fontshape{#5}%
  \selectfont}%
\fi\endgroup%
{\renewcommand{\dashlinestretch}{30}
\begin{picture}(2486,2209)(0,-10)
\thicklines
\path(459,1057)(593,1069)
\path(469,967)(604,984)
\path(611,1583)(476,1583)
\path(606,611)(471,611)
\path(536,677)(536,541)
\path(1834,621)(1969,621)
\path(1839,1573)(1974,1573)
\path(1909,1507)(1909,1643)
\thinlines
\put(1225,1102){\ellipse{1400}{1400}}
\path(2044,1829)(1963,1828)(1965,1908)
\path(2172,2035)(2109,1971)
\path(2046,1907)(1981,1844)
\path(1918,1780)(1854,1717)
\path(2300,2162)(2236,2098)
\path(1791,1652)(1727,1589)
\path(401,355)(482,356)(480,276)
\path(273,149)(336,213)
\path(399,277)(464,340)
\path(527,404)(591,467)
\path(145,22)(209,86)
\path(654,532)(718,595)
\path(406,1839)(487,1838)(485,1918)
\path(278,2045)(341,1981)
\path(404,1917)(469,1854)
\path(532,1790)(596,1727)
\path(150,2172)(214,2108)
\path(659,1662)(723,1599)
\path(2039,365)(1958,366)(1960,286)
\path(2167,159)(2104,223)
\path(2041,287)(1976,350)
\path(1913,414)(1849,477)
\path(2295,32)(2231,96)
\path(1786,542)(1722,605)
\path(589,1178)(547,1253)(466,1202)
\put(5,287){\makebox(0,0)[lb]{\smash{{{\SetFigFont{10}{12.0}{\rmdefault}{\mddefault}{\itdefault}p}}}}}
\put(2305,289){\makebox(0,0)[lb]{\smash{{{\SetFigFont{10}{12.0}{\rmdefault}{\mddefault}{\itdefault}p}}}}}
\put(2255,1869){\makebox(0,0)[lb]{\smash{{{\SetFigFont{10}{12.0}{\rmdefault}{\mddefault}{\itdefault}p}}}}}
\put(0,1869){\makebox(0,0)[lb]{\smash{{{\SetFigFont{10}{12.0}{\rmdefault}{\mddefault}{\itdefault}p}}}}}
\put(110,1764){\makebox(0,0)[lb]{\smash{{{\SetFigFont{5}{6.0}{\rmdefault}{\mddefault}{\itdefault}3}}}}}
\put(105,177){\makebox(0,0)[lb]{\smash{{{\SetFigFont{5}{6.0}{\rmdefault}{\mddefault}{\itdefault}2}}}}}
\put(2375,1774){\makebox(0,0)[lb]{\smash{{{\SetFigFont{5}{6.0}{\rmdefault}{\mddefault}{\itdefault}4}}}}}
\put(2410,179){\makebox(0,0)[lb]{\smash{{{\SetFigFont{5}{6.0}{\rmdefault}{\mddefault}{\itdefault}1}}}}}
\end{picture}
}
  \end{array}\; + \;   
\nonumber \\ {} & {} & {}  \nonumber \\
  \begin{array}[h]{c} 
\setlength{\unitlength}{0.00041667in}
\begingroup\makeatletter\ifx\SetFigFont\undefined%
\gdef\SetFigFont#1#2#3#4#5{%
  \reset@font\fontsize{#1}{#2pt}%
  \fontfamily{#3}\fontseries{#4}\fontshape{#5}%
  \selectfont}%
\fi\endgroup%
{\renewcommand{\dashlinestretch}{30}
\begin{picture}(2486,2209)(0,-10)
\thicklines
\path(459,1057)(593,1069)
\path(469,967)(604,984)
\path(611,1583)(476,1583)
\path(541,1517)(541,1653)
\path(606,611)(471,611)
\path(1834,621)(1969,621)
\path(1904,687)(1904,551)
\path(1839,1573)(1974,1573)
\path(1909,1507)(1909,1643)
\thinlines
\put(1225,1102){\ellipse{1400}{1400}}
\path(2044,1829)(1963,1828)(1965,1908)
\path(2172,2035)(2109,1971)
\path(2046,1907)(1981,1844)
\path(1918,1780)(1854,1717)
\path(2300,2162)(2236,2098)
\path(1791,1652)(1727,1589)
\path(401,355)(482,356)(480,276)
\path(273,149)(336,213)
\path(399,277)(464,340)
\path(527,404)(591,467)
\path(145,22)(209,86)
\path(654,532)(718,595)
\path(406,1839)(487,1838)(485,1918)
\path(278,2045)(341,1981)
\path(404,1917)(469,1854)
\path(532,1790)(596,1727)
\path(150,2172)(214,2108)
\path(659,1662)(723,1599)
\path(2039,365)(1958,366)(1960,286)
\path(2167,159)(2104,223)
\path(2041,287)(1976,350)
\path(1913,414)(1849,477)
\path(2295,32)(2231,96)
\path(1786,542)(1722,605)
\path(589,1178)(547,1253)(466,1202)
\put(5,287){\makebox(0,0)[lb]{\smash{{{\SetFigFont{10}{12.0}{\rmdefault}{\mddefault}{\itdefault}p}}}}}
\put(2305,289){\makebox(0,0)[lb]{\smash{{{\SetFigFont{10}{12.0}{\rmdefault}{\mddefault}{\itdefault}p}}}}}
\put(2255,1869){\makebox(0,0)[lb]{\smash{{{\SetFigFont{10}{12.0}{\rmdefault}{\mddefault}{\itdefault}p}}}}}
\put(0,1869){\makebox(0,0)[lb]{\smash{{{\SetFigFont{10}{12.0}{\rmdefault}{\mddefault}{\itdefault}p}}}}}
\put(110,1764){\makebox(0,0)[lb]{\smash{{{\SetFigFont{5}{6.0}{\rmdefault}{\mddefault}{\itdefault}3}}}}}
\put(105,177){\makebox(0,0)[lb]{\smash{{{\SetFigFont{5}{6.0}{\rmdefault}{\mddefault}{\itdefault}2}}}}}
\put(2375,1774){\makebox(0,0)[lb]{\smash{{{\SetFigFont{5}{6.0}{\rmdefault}{\mddefault}{\itdefault}4}}}}}
\put(2410,179){\makebox(0,0)[lb]{\smash{{{\SetFigFont{5}{6.0}{\rmdefault}{\mddefault}{\itdefault}1}}}}}
\end{picture}
}
  \end{array}\; +\; 
&
  \begin{array}[h]{c}
\setlength{\unitlength}{0.00041667in}
\begingroup\makeatletter\ifx\SetFigFont\undefined%
\gdef\SetFigFont#1#2#3#4#5{%
  \reset@font\fontsize{#1}{#2pt}%
  \fontfamily{#3}\fontseries{#4}\fontshape{#5}%
  \selectfont}%
\fi\endgroup%
{\renewcommand{\dashlinestretch}{30}
\begin{picture}(2486,2209)(0,-10)
\thicklines
\path(459,1057)(593,1069)
\path(469,967)(604,984)
\path(611,1583)(476,1583)
\path(606,611)(471,611)
\path(1834,621)(1969,621)
\path(1904,687)(1904,551)
\path(1839,1573)(1974,1573)
\path(1909,1507)(1909,1643)
\thinlines
\put(1225,1102){\ellipse{1400}{1400}}
\path(2044,1829)(1963,1828)(1965,1908)
\path(2172,2035)(2109,1971)
\path(2046,1907)(1981,1844)
\path(1918,1780)(1854,1717)
\path(2300,2162)(2236,2098)
\path(1791,1652)(1727,1589)
\path(401,355)(482,356)(480,276)
\path(273,149)(336,213)
\path(399,277)(464,340)
\path(527,404)(591,467)
\path(145,22)(209,86)
\path(654,532)(718,595)
\path(406,1839)(487,1838)(485,1918)
\path(278,2045)(341,1981)
\path(404,1917)(469,1854)
\path(532,1790)(596,1727)
\path(150,2172)(214,2108)
\path(659,1662)(723,1599)
\path(2039,365)(1958,366)(1960,286)
\path(2167,159)(2104,223)
\path(2041,287)(1976,350)
\path(1913,414)(1849,477)
\path(2295,32)(2231,96)
\path(1786,542)(1722,605)
\path(589,1178)(547,1253)(466,1202)
\put(5,287){\makebox(0,0)[lb]{\smash{{{\SetFigFont{10}{12.0}{\rmdefault}{\mddefault}{\itdefault}p}}}}}
\put(2305,289){\makebox(0,0)[lb]{\smash{{{\SetFigFont{10}{12.0}{\rmdefault}{\mddefault}{\itdefault}p}}}}}
\put(2255,1869){\makebox(0,0)[lb]{\smash{{{\SetFigFont{10}{12.0}{\rmdefault}{\mddefault}{\itdefault}p}}}}}
\put(0,1869){\makebox(0,0)[lb]{\smash{{{\SetFigFont{10}{12.0}{\rmdefault}{\mddefault}{\itdefault}p}}}}}
\put(110,1764){\makebox(0,0)[lb]{\smash{{{\SetFigFont{5}{6.0}{\rmdefault}{\mddefault}{\itdefault}3}}}}}
\put(105,177){\makebox(0,0)[lb]{\smash{{{\SetFigFont{5}{6.0}{\rmdefault}{\mddefault}{\itdefault}2}}}}}
\put(2375,1774){\makebox(0,0)[lb]{\smash{{{\SetFigFont{5}{6.0}{\rmdefault}{\mddefault}{\itdefault}4}}}}}
\put(2410,179){\makebox(0,0)[lb]{\smash{{{\SetFigFont{5}{6.0}{\rmdefault}{\mddefault}{\itdefault}1}}}}}
\end{picture}
}
  \end{array}
\; + \;    
  \begin{array}[h]{c}
\setlength{\unitlength}{0.00041667in}
\begingroup\makeatletter\ifx\SetFigFont\undefined%
\gdef\SetFigFont#1#2#3#4#5{%
  \reset@font\fontsize{#1}{#2pt}%
  \fontfamily{#3}\fontseries{#4}\fontshape{#5}%
  \selectfont}%
\fi\endgroup%
{\renewcommand{\dashlinestretch}{30}
\begin{picture}(2486,2209)(0,-10)
\thicklines
\path(459,1057)(593,1069)
\path(469,967)(604,984)
\path(611,1583)(476,1583)
\path(541,1517)(541,1653)
\path(606,611)(471,611)
\path(1834,621)(1969,621)
\path(1839,1573)(1974,1573)
\path(1909,1507)(1909,1643)
\thinlines
\put(1225,1102){\ellipse{1400}{1400}}
\path(2044,1829)(1963,1828)(1965,1908)
\path(2172,2035)(2109,1971)
\path(2046,1907)(1981,1844)
\path(1918,1780)(1854,1717)
\path(2300,2162)(2236,2098)
\path(1791,1652)(1727,1589)
\path(401,355)(482,356)(480,276)
\path(273,149)(336,213)
\path(399,277)(464,340)
\path(527,404)(591,467)
\path(145,22)(209,86)
\path(654,532)(718,595)
\path(406,1839)(487,1838)(485,1918)
\path(278,2045)(341,1981)
\path(404,1917)(469,1854)
\path(532,1790)(596,1727)
\path(150,2172)(214,2108)
\path(659,1662)(723,1599)
\path(2039,365)(1958,366)(1960,286)
\path(2167,159)(2104,223)
\path(2041,287)(1976,350)
\path(1913,414)(1849,477)
\path(2295,32)(2231,96)
\path(1786,542)(1722,605)
\path(589,1178)(547,1253)(466,1202)
\put(5,287){\makebox(0,0)[lb]{\smash{{{\SetFigFont{10}{12.0}{\rmdefault}{\mddefault}{\itdefault}p}}}}}
\put(2305,289){\makebox(0,0)[lb]{\smash{{{\SetFigFont{10}{12.0}{\rmdefault}{\mddefault}{\itdefault}p}}}}}
\put(2255,1869){\makebox(0,0)[lb]{\smash{{{\SetFigFont{10}{12.0}{\rmdefault}{\mddefault}{\itdefault}p}}}}}
\put(0,1869){\makebox(0,0)[lb]{\smash{{{\SetFigFont{10}{12.0}{\rmdefault}{\mddefault}{\itdefault}p}}}}}
\put(110,1764){\makebox(0,0)[lb]{\smash{{{\SetFigFont{5}{6.0}{\rmdefault}{\mddefault}{\itdefault}3}}}}}
\put(105,177){\makebox(0,0)[lb]{\smash{{{\SetFigFont{5}{6.0}{\rmdefault}{\mddefault}{\itdefault}2}}}}}
\put(2375,1774){\makebox(0,0)[lb]{\smash{{{\SetFigFont{5}{6.0}{\rmdefault}{\mddefault}{\itdefault}4}}}}}
\put(2410,179){\makebox(0,0)[lb]{\smash{{{\SetFigFont{5}{6.0}{\rmdefault}{\mddefault}{\itdefault}1}}}}}
\end{picture}
}
  \end{array}
\; +\; 
&
  \begin{array}[h]{c}
\setlength{\unitlength}{0.00041667in}
\begingroup\makeatletter\ifx\SetFigFont\undefined%
\gdef\SetFigFont#1#2#3#4#5{%
  \reset@font\fontsize{#1}{#2pt}%
  \fontfamily{#3}\fontseries{#4}\fontshape{#5}%
  \selectfont}%
\fi\endgroup%
{\renewcommand{\dashlinestretch}{30}
\begin{picture}(2486,2209)(0,-10)
\thicklines
\path(459,1057)(593,1069)
\path(469,967)(604,984)
\path(611,1583)(476,1583)
\path(606,611)(471,611)
\path(1834,621)(1969,621)
\path(1839,1573)(1974,1573)
\path(1909,1507)(1909,1643)
\thinlines
\put(1225,1102){\ellipse{1400}{1400}}
\path(2044,1829)(1963,1828)(1965,1908)
\path(2172,2035)(2109,1971)
\path(2046,1907)(1981,1844)
\path(1918,1780)(1854,1717)
\path(2300,2162)(2236,2098)
\path(1791,1652)(1727,1589)
\path(401,355)(482,356)(480,276)
\path(273,149)(336,213)
\path(399,277)(464,340)
\path(527,404)(591,467)
\path(145,22)(209,86)
\path(654,532)(718,595)
\path(406,1839)(487,1838)(485,1918)
\path(278,2045)(341,1981)
\path(404,1917)(469,1854)
\path(532,1790)(596,1727)
\path(150,2172)(214,2108)
\path(659,1662)(723,1599)
\path(2039,365)(1958,366)(1960,286)
\path(2167,159)(2104,223)
\path(2041,287)(1976,350)
\path(1913,414)(1849,477)
\path(2295,32)(2231,96)
\path(1786,542)(1722,605)
\path(589,1178)(547,1253)(466,1202)
\put(5,287){\makebox(0,0)[lb]{\smash{{{\SetFigFont{10}{12.0}{\rmdefault}{\mddefault}{\itdefault}p}}}}}
\put(2305,289){\makebox(0,0)[lb]{\smash{{{\SetFigFont{10}{12.0}{\rmdefault}{\mddefault}{\itdefault}p}}}}}
\put(2255,1869){\makebox(0,0)[lb]{\smash{{{\SetFigFont{10}{12.0}{\rmdefault}{\mddefault}{\itdefault}p}}}}}
\put(0,1869){\makebox(0,0)[lb]{\smash{{{\SetFigFont{10}{12.0}{\rmdefault}{\mddefault}{\itdefault}p}}}}}
\put(110,1764){\makebox(0,0)[lb]{\smash{{{\SetFigFont{5}{6.0}{\rmdefault}{\mddefault}{\itdefault}3}}}}}
\put(105,177){\makebox(0,0)[lb]{\smash{{{\SetFigFont{5}{6.0}{\rmdefault}{\mddefault}{\itdefault}2}}}}}
\put(2375,1774){\makebox(0,0)[lb]{\smash{{{\SetFigFont{5}{6.0}{\rmdefault}{\mddefault}{\itdefault}4}}}}}
\put(2410,179){\makebox(0,0)[lb]{\smash{{{\SetFigFont{5}{6.0}{\rmdefault}{\mddefault}{\itdefault}1}}}}}
\end{picture}
}
  \end{array}    
\nonumber \\  {} & {} & {} \nonumber \\
&  
\; = \;   \begin{array}[h]{c}
\setlength{\unitlength}{0.00041667in}
\begingroup\makeatletter\ifx\SetFigFont\undefined%
\gdef\SetFigFont#1#2#3#4#5{%
  \reset@font\fontsize{#1}{#2pt}%
  \fontfamily{#3}\fontseries{#4}\fontshape{#5}%
  \selectfont}%
\fi\endgroup%
{\renewcommand{\dashlinestretch}{30}
\begin{picture}(2486,2209)(0,-10)
\thicklines
\path(459,1057)(593,1069)
\path(469,967)(604,984)
\put(2010,949){\makebox(0,0)[lb]{\smash{{{\SetFigFont{10}{12.0}{\rmdefault}{\mddefault}{\itdefault}R}}}}}
\put(1105,104){\makebox(0,0)[lb]{\smash{{{\SetFigFont{10}{12.0}{\rmdefault}{\mddefault}{\itdefault}R}}}}}
\put(1130,1894){\makebox(0,0)[lb]{\smash{{{\SetFigFont{10}{12.0}{\rmdefault}{\mddefault}{\itdefault}A}}}}}
\thinlines
\put(1225,1102){\ellipse{1400}{1400}}
\path(2044,1829)(1963,1828)(1965,1908)
\path(2172,2035)(2109,1971)
\path(2046,1907)(1981,1844)
\path(1918,1780)(1854,1717)
\path(2300,2162)(2236,2098)
\path(1791,1652)(1727,1589)
\path(401,355)(482,356)(480,276)
\path(273,149)(336,213)
\path(399,277)(464,340)
\path(527,404)(591,467)
\path(145,22)(209,86)
\path(654,532)(718,595)
\path(406,1839)(487,1838)(485,1918)
\path(278,2045)(341,1981)
\path(404,1917)(469,1854)
\path(532,1790)(596,1727)
\path(150,2172)(214,2108)
\path(659,1662)(723,1599)
\path(2039,365)(1958,366)(1960,286)
\path(2167,159)(2104,223)
\path(2041,287)(1976,350)
\path(1913,414)(1849,477)
\path(2295,32)(2231,96)
\path(1786,542)(1722,605)
\path(589,1178)(547,1253)(466,1202)
\put(5,287){\makebox(0,0)[lb]{\smash{{{\SetFigFont{10}{12.0}{\rmdefault}{\mddefault}{\itdefault}p}}}}}
\put(2305,289){\makebox(0,0)[lb]{\smash{{{\SetFigFont{10}{12.0}{\rmdefault}{\mddefault}{\itdefault}p}}}}}
\put(2255,1869){\makebox(0,0)[lb]{\smash{{{\SetFigFont{10}{12.0}{\rmdefault}{\mddefault}{\itdefault}p}}}}}
\put(0,1869){\makebox(0,0)[lb]{\smash{{{\SetFigFont{10}{12.0}{\rmdefault}{\mddefault}{\itdefault}p}}}}}
\put(110,1764){\makebox(0,0)[lb]{\smash{{{\SetFigFont{5}{6.0}{\rmdefault}{\mddefault}{\itdefault}3}}}}}
\put(105,177){\makebox(0,0)[lb]{\smash{{{\SetFigFont{5}{6.0}{\rmdefault}{\mddefault}{\itdefault}2}}}}}
\put(2375,1774){\makebox(0,0)[lb]{\smash{{{\SetFigFont{5}{6.0}{\rmdefault}{\mddefault}{\itdefault}4}}}}}
\put(2410,179){\makebox(0,0)[lb]{\smash{{{\SetFigFont{5}{6.0}{\rmdefault}{\mddefault}{\itdefault}1}}}}}
\end{picture}
} \;\;\;\;\;\;\;\;\;\;
  \end{array} 
& .
\end{eqnarray} 
The right-hand side of this equation can be represented
diagrammatically as the forward scattering amplitude shown
in Fig. (4c).
\begin{figure}[h!]
\[
\begin{array}{c}
\begin{array}{ccc}
\setlength{\unitlength}{0.00035in}
\begingroup\makeatletter\ifx\SetFigFont\undefined%
\gdef\SetFigFont#1#2#3#4#5{%
  \reset@font\fontsize{#1}{#2pt}%
  \fontfamily{#3}\fontseries{#4}\fontshape{#5}%
  \selectfont}%
\fi\endgroup%
{\renewcommand{\dashlinestretch}{30}
\begin{picture}(4201,2204)(0,-10)
\thicklines
\path(447,428)(532,343)
\path(391,372)(476,287)
\path(3786,293)(3873,379)
\path(3844,237)(3928,322)
\thinlines

\path(619,432)(620,513)(540,512)
\path(3786,293)(3873,379)
\path(3844,237)(3928,322)
\path(3738,510)(3739,430)(3658,431)
\put(1410,2200){\makebox(0,0)[lb]{\smash{{{\SetFigFont{8}{9.6}{\rmdefault}{\mddefault}{\itdefault}p}}}}}
\put(1500,2030){\makebox(0,0)[lb]{\smash{{{\SetFigFont{5}{6.0}{\rmdefault}{\mddefault}{\itdefault}2}}}}}
\put(0,1950){\makebox(0,0)[lb]{\smash{{{\SetFigFont{8}{9.6}{\rmdefault}{\mddefault}{\itdefault}p}}}}}
\put(80,1780){\makebox(0,0)[lb]{\smash{{{\SetFigFont{5}{6.0}{\rmdefault}{\mddefault}{\itdefault}1}}}}}
\put(2400,2200){\makebox(0,0)[lb]{\smash{{{\SetFigFont{8}{9.6}{\rmdefault}{\mddefault}{\itdefault}p}}}}}
\put(2495,2030){\makebox(0,0)[lb]{\smash{{{\SetFigFont{5}{6.0}{\rmdefault}{\mddefault}{\itdefault}3}}}}}
\put(4035,1950){\makebox(0,0)[lb]{\smash{{{\SetFigFont{8}{9.6}{\rmdefault}{\mddefault}{\itdefault}p}}}}}
\put(4125,1780){\makebox(0,0)[lb]{\smash{{{\SetFigFont{5}{6.0}{\rmdefault}{\mddefault}{\itdefault}4}}}}}
\path(3917,1411)(3853,1348)
\path(3789,1285)(3726,1220)
\path(3662,1157)(3599,1093)
\path(4044,1539)(3980,1475)
\path(3534,1030)(3471,966)
\path(3408,902)(3344,839)
\path(4171,1666)(4108,1602)
\path(355,1408)(418,1344)
\path(481,1280)(546,1217)
\path(609,1153)(673,1090)
\path(227,1535)(291,1471)
\path(736,1025)(800,962)
\path(864,899)(927,835)
\path(100,1662)(164,1599)
\path(1596,1660)(1596,1570)
\path(1596,1480)(1596,1390)
\path(1596,1300)(1596,1210)
\path(1596,1840)(1596,1750)
\path(1596,1120)(1596,1030)
\path(1596,940)(1596,850)
\path(2569,1660)(2569,1570)
\path(2569,1480)(2569,1390)
\path(2569,1300)(2569,1210)
\path(2569,1840)(2569,1750)
\path(2569,1120)(2569,1030)
\path(2569,940)(2569,850)
\path(2045,780)(2103,837)(2045,893)
\path(134,31)(939,836)(3336,836)
	(4135,37)(4135,44)
\path(1224,789)(1282,846)(1224,902)
\path(2946,786)(3004,843)(2946,899)
\path(2622,1430)(2566,1372)(2510,1430)
\path(1650,1427)(1594,1369)(1538,1427)
\path(483,1202)(564,1201)(562,1281)
\path(3787,1204)(3706,1203)(3708,1283)
\put(0,300){\makebox(0,0)[lb]{\smash{{{\SetFigFont{8}{9.6}{\rmdefault}{\mddefault}{\itdefault}k}}}}}
\put(4035,300){\makebox(0,0)[lb]{\smash{{{\SetFigFont{8}{9.6}{\rmdefault}{\mddefault}{\itdefault}k}}}}}
\put(1100,1010){\makebox(0,0)[lb]{\smash{{{\SetFigFont{8}{9.6}{\rmdefault}{\mddefault}{\itdefault}A}}}}}
\put(1900,-300){\makebox(0,0)[lb]{\smash{{{\SetFigFont{8}{9.6}{\rmdefault}{\mddefault}{\rmdefault}(a)}}}}}
\put(2000,1010){\makebox(0,0)[lb]{\smash{{{\SetFigFont{8}{9.6}{\rmdefault}{\mddefault}{\itdefault}A}}}}}
\put(2900,1010){\makebox(0,0)[lb]{\smash{{{\SetFigFont{8}{9.6}{\rmdefault}{\mddefault}{\itdefault}A}}}}}
\end{picture}}
\; \; \; \; 
&
\setlength{\unitlength}{0.00035in}
\begingroup\makeatletter\ifx\SetFigFont\undefined%
\gdef\SetFigFont#1#2#3#4#5{%
  \reset@font\fontsize{#1}{#2pt}%
  \fontfamily{#3}\fontseries{#4}\fontshape{#5}%
  \selectfont}%
\fi\endgroup%
{\renewcommand{\dashlinestretch}{30}
\begin{picture}(4201,2204)(0,-10)
\thicklines
\path(447,428)(532,343)
\path(391,372)(476,287)
\path(3786,293)(3873,379)
\path(3844,237)(3928,322)
\thinlines
\path(619,432)(620,513)(540,512)
\path(3738,510)(3739,430)(3658,431)
\put(1410,2200){\makebox(0,0)[lb]{\smash{{{\SetFigFont{8}{9.6}{\rmdefault}{\mddefault}{\itdefault}p}}}}}
\put(1500,2030){\makebox(0,0)[lb]{\smash{{{\SetFigFont{5}{6.0}{\rmdefault}{\mddefault}{\itdefault}3}}}}}
\put(0,1950){\makebox(0,0)[lb]{\smash{{{\SetFigFont{8}{9.6}{\rmdefault}{\mddefault}{\itdefault}p}}}}}
\put(80,1780){\makebox(0,0)[lb]{\smash{{{\SetFigFont{5}{6.0}{\rmdefault}{\mddefault}{\itdefault}2}}}}}
\put(2400,2200){\makebox(0,0)[lb]{\smash{{{\SetFigFont{8}{9.6}{\rmdefault}{\mddefault}{\itdefault}p}}}}}
\put(2495,2030){\makebox(0,0)[lb]{\smash{{{\SetFigFont{5}{6.0}{\rmdefault}{\mddefault}{\itdefault}4}}}}}
\put(4035,1950){\makebox(0,0)[lb]{\smash{{{\SetFigFont{8}{9.6}{\rmdefault}{\mddefault}{\itdefault}p}}}}}
\put(4125,1780){\makebox(0,0)[lb]{\smash{{{\SetFigFont{5}{6.0}{\rmdefault}{\mddefault}{\itdefault}1}}}}}
\path(3917,1411)(3853,1348)
\path(3789,1285)(3726,1220)
\path(3662,1157)(3599,1093)
\path(4044,1539)(3980,1475)
\path(3534,1030)(3471,966)
\path(3408,902)(3344,839)
\path(4171,1666)(4108,1602)
\path(355,1408)(418,1344)
\path(481,1280)(546,1217)
\path(609,1153)(673,1090)
\path(227,1535)(291,1471)
\path(736,1025)(800,962)
\path(864,899)(927,835)
\path(100,1662)(164,1599)
\path(1596,1660)(1596,1570)
\path(1596,1480)(1596,1390)
\path(1596,1300)(1596,1210)
\path(1596,1840)(1596,1750)
\path(1596,1120)(1596,1030)
\path(1596,940)(1596,850)
\path(2569,1660)(2569,1570)
\path(2569,1480)(2569,1390)
\path(2569,1300)(2569,1210)
\path(2569,1840)(2569,1750)
\path(2569,1120)(2569,1030)
\path(2569,940)(2569,850)
\path(2045,780)(2103,837)(2045,893)
\path(134,31)(939,836)(3336,836)
	(4135,37)(4135,44)
\path(1224,789)(1282,846)(1224,902)
\path(2946,786)(3004,843)(2946,899)
\path(2622,1430)(2566,1372)(2510,1430)
\path(1650,1427)(1594,1369)(1538,1427)
\path(483,1202)(564,1201)(562,1281)
\path(3787,1204)(3706,1203)(3708,1283)
\put(0,300){\makebox(0,0)[lb]{\smash{{{\SetFigFont{8}{9.6}{\rmdefault}{\mddefault}{\itdefault}k}}}}}
\put(4035,300){\makebox(0,0)[lb]{\smash{{{\SetFigFont{8}{9.6}{\rmdefault}{\mddefault}{\itdefault}k}}}}}
\put(1100,1010){\makebox(0,0)[lb]{\smash{{{\SetFigFont{8}{9.6}{\rmdefault}{\mddefault}{\itdefault}A}}}}}
\put(1900,-300){\makebox(0,0)[lb]{\smash{{{\SetFigFont{8}{9.6}{\rmdefault}{\mddefault}{\rmdefault}(b)}}}}}
\put(2000,1010){\makebox(0,0)[lb]{\smash{{{\SetFigFont{8}{9.6}{\rmdefault}{\mddefault}{\itdefault}A}}}}}
\put(2900,1010){\makebox(0,0)[lb]{\smash{{{\SetFigFont{8}{9.6}{\rmdefault}{\mddefault}{\itdefault}R}}}}}
\end{picture}}
\\ {}  \\ 
\setlength{\unitlength}{0.00035in}
\begingroup\makeatletter\ifx\SetFigFont\undefined%
\gdef\SetFigFont#1#2#3#4#5{%
  \reset@font\fontsize{#1}{#2pt}%
  \fontfamily{#3}\fontseries{#4}\fontshape{#5}%
  \selectfont}%
\fi\endgroup%
{\renewcommand{\dashlinestretch}{30}
\begin{picture}(4201,2204)(0,-10)
\thicklines
\path(447,428)(532,343)
\path(391,372)(476,287)
\path(3786,293)(3873,379)
\path(3844,237)(3928,322)
\thinlines
\path(619,432)(620,513)(540,512)
\path(3738,510)(3739,430)(3658,431)
\put(1410,2200){\makebox(0,0)[lb]{\smash{{{\SetFigFont{8}{9.6}{\rmdefault}{\mddefault}{\itdefault}p}}}}}
\put(1500,2030){\makebox(0,0)[lb]{\smash{{{\SetFigFont{5}{6.0}{\rmdefault}{\mddefault}{\itdefault}4}}}}}
\put(0,1950){\makebox(0,0)[lb]{\smash{{{\SetFigFont{8}{9.6}{\rmdefault}{\mddefault}{\itdefault}p}}}}}
\put(80,1780){\makebox(0,0)[lb]{\smash{{{\SetFigFont{5}{6.0}{\rmdefault}{\mddefault}{\itdefault}3}}}}}
\put(2400,2200){\makebox(0,0)[lb]{\smash{{{\SetFigFont{8}{9.6}{\rmdefault}{\mddefault}{\itdefault}p}}}}}
\put(2495,2030){\makebox(0,0)[lb]{\smash{{{\SetFigFont{5}{6.0}{\rmdefault}{\mddefault}{\itdefault}1}}}}}
\put(4035,1950){\makebox(0,0)[lb]{\smash{{{\SetFigFont{8}{9.6}{\rmdefault}{\mddefault}{\itdefault}p}}}}}
\put(4125,1780){\makebox(0,0)[lb]{\smash{{{\SetFigFont{5}{6.0}{\rmdefault}{\mddefault}{\itdefault}2}}}}}
\path(3917,1411)(3853,1348)
\path(3789,1285)(3726,1220)
\path(3662,1157)(3599,1093)
\path(4044,1539)(3980,1475)
\path(3534,1030)(3471,966)
\path(3408,902)(3344,839)
\path(4171,1666)(4108,1602)
\path(355,1408)(418,1344)
\path(481,1280)(546,1217)
\path(609,1153)(673,1090)
\path(227,1535)(291,1471)
\path(736,1025)(800,962)
\path(864,899)(927,835)
\path(100,1662)(164,1599)
\path(1596,1660)(1596,1570)
\path(1596,1480)(1596,1390)
\path(1596,1300)(1596,1210)
\path(1596,1840)(1596,1750)
\path(1596,1120)(1596,1030)
\path(1596,940)(1596,850)
\path(2569,1660)(2569,1570)
\path(2569,1480)(2569,1390)
\path(2569,1300)(2569,1210)
\path(2569,1840)(2569,1750)
\path(2569,1120)(2569,1030)
\path(2569,940)(2569,850)
\path(2045,780)(2103,837)(2045,893)
\path(134,31)(939,836)(3336,836)
	(4135,37)(4135,44)
\path(1224,789)(1282,846)(1224,902)
\path(2946,786)(3004,843)(2946,899)
\path(2622,1430)(2566,1372)(2510,1430)
\path(1650,1427)(1594,1369)(1538,1427)
\path(483,1202)(564,1201)(562,1281)
\path(3787,1204)(3706,1203)(3708,1283)
\put(0,300){\makebox(0,0)[lb]{\smash{{{\SetFigFont{8}{9.6}{\rmdefault}{\mddefault}{\itdefault}k}}}}}
\put(4035,300){\makebox(0,0)[lb]{\smash{{{\SetFigFont{8}{9.6}{\rmdefault}{\mddefault}{\itdefault}k}}}}}
\put(1100,1010){\makebox(0,0)[lb]{\smash{{{\SetFigFont{8}{9.6}{\rmdefault}{\mddefault}{\itdefault}A}}}}}
\put(1900,-300){\makebox(0,0)[lb]{\smash{{{\SetFigFont{8}{9.6}{\rmdefault}{\mddefault}{\rmdefault}(c)}}}}}
\put(2000,1010){\makebox(0,0)[lb]{\smash{{{\SetFigFont{8}{9.6}{\rmdefault}{\mddefault}{\itdefault}R}}}}}
\put(2900,1010){\makebox(0,0)[lb]{\smash{{{\SetFigFont{8}{9.6}{\rmdefault}{\mddefault}{\itdefault}R}}}}}
\end{picture}} 
\; \; \; \; 
&
\setlength{\unitlength}{0.00035in}
\begingroup\makeatletter\ifx\SetFigFont\undefined%
\gdef\SetFigFont#1#2#3#4#5{%
  \reset@font\fontsize{#1}{#2pt}%
  \fontfamily{#3}\fontseries{#4}\fontshape{#5}%
  \selectfont}%
\fi\endgroup%
{\renewcommand{\dashlinestretch}{30}
\begin{picture}(4201,2204)(0,-10)
\thicklines
\path(447,428)(532,343)
\path(391,372)(476,287)
\path(3786,293)(3873,379)
\path(3844,237)(3928,322)
\thinlines
\path(619,432)(620,513)(540,512)
\path(3786,293)(3873,379)
\path(3844,237)(3928,322)
\path(3738,510)(3739,430)(3658,431)
\put(1410,2200){\makebox(0,0)[lb]{\smash{{{\SetFigFont{8}{9.6}{\rmdefault}{\mddefault}{\itdefault}p}}}}}
\put(1500,2030){\makebox(0,0)[lb]{\smash{{{\SetFigFont{5}{6.0}{\rmdefault}{\mddefault}{\itdefault}1}}}}}
\put(0,1950){\makebox(0,0)[lb]{\smash{{{\SetFigFont{8}{9.6}{\rmdefault}{\mddefault}{\itdefault}p}}}}}
\put(80,1780){\makebox(0,0)[lb]{\smash{{{\SetFigFont{5}{6.0}{\rmdefault}{\mddefault}{\itdefault}4}}}}}
\put(2400,2200){\makebox(0,0)[lb]{\smash{{{\SetFigFont{8}{9.6}{\rmdefault}{\mddefault}{\itdefault}p}}}}}
\put(2495,2030){\makebox(0,0)[lb]{\smash{{{\SetFigFont{5}{6.0}{\rmdefault}{\mddefault}{\itdefault}2}}}}}
\put(4035,1950){\makebox(0,0)[lb]{\smash{{{\SetFigFont{8}{9.6}{\rmdefault}{\mddefault}{\itdefault}p}}}}}
\put(4125,1780){\makebox(0,0)[lb]{\smash{{{\SetFigFont{5}{6.0}{\rmdefault}{\mddefault}{\itdefault}3}}}}}
\path(3917,1411)(3853,1348)
\path(3789,1285)(3726,1220)
\path(3662,1157)(3599,1093)
\path(4044,1539)(3980,1475)
\path(3534,1030)(3471,966)
\path(3408,902)(3344,839)
\path(4171,1666)(4108,1602)
\path(355,1408)(418,1344)
\path(481,1280)(546,1217)
\path(609,1153)(673,1090)
\path(227,1535)(291,1471)
\path(736,1025)(800,962)
\path(864,899)(927,835)
\path(100,1662)(164,1599)
\path(1596,1660)(1596,1570)
\path(1596,1480)(1596,1390)
\path(1596,1300)(1596,1210)
\path(1596,1840)(1596,1750)
\path(1596,1120)(1596,1030)
\path(1596,940)(1596,850)
\path(2569,1660)(2569,1570)
\path(2569,1480)(2569,1390)
\path(2569,1300)(2569,1210)
\path(2569,1840)(2569,1750)
\path(2569,1120)(2569,1030)
\path(2569,940)(2569,850)
\path(2045,780)(2103,837)(2045,893)
\path(134,31)(939,836)(3336,836)
	(4135,37)(4135,44)
\path(1224,789)(1282,846)(1224,902)
\path(2946,786)(3004,843)(2946,899)
\path(2622,1430)(2566,1372)(2510,1430)
\path(1650,1427)(1594,1369)(1538,1427)
\path(483,1202)(564,1201)(562,1281)
\path(3787,1204)(3706,1203)(3708,1283)
\put(1900,-300){\makebox(0,0)[lb]{\smash{{{\SetFigFont{8}{9.6}{\rmdefault}{\mddefault}{\rmdefault}(d)}}}}}
\put(0,300){\makebox(0,0)[lb]{\smash{{{\SetFigFont{8}{9.6}{\rmdefault}{\mddefault}{\itdefault}k}}}}}
\put(4035,300){\makebox(0,0)[lb]{\smash{{{\SetFigFont{8}{9.6}{\rmdefault}{\mddefault}{\itdefault}k}}}}}
\put(1100,1010){\makebox(0,0)[lb]{\smash{{{\SetFigFont{8}{9.6}{\rmdefault}{\mddefault}{\itdefault}R}}}}}
\put(2000,1010){\makebox(0,0)[lb]{\smash{{{\SetFigFont{8}{9.6}{\rmdefault}{\mddefault}{\itdefault}R}}}}}
\put(2900,1010){\makebox(0,0)[lb]{\smash{{{\SetFigFont{8}{9.6}{\rmdefault}{\mddefault}{\itdefault}R}}}}}
\end{picture}
}

\end{array}
\end{array}
\]
   \label{fig3}
\bigskip
\caption{Forward scattering amplitudes associated with the 
retarded 4-point function. The internal on-shell momentum $k$ 
is integrated over with an appropriate statistical
factor.}
  \end{figure}
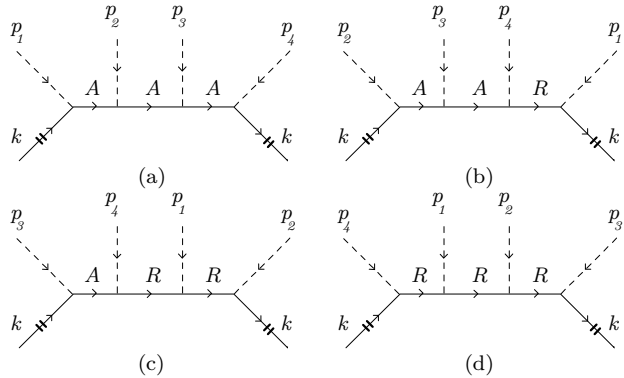
Thus,  this set of  diagrams uniquely  gives  the retarded four  point
amplitude as 
\begin{equation}
\Gamma_{R}^{(4)}  =  \Gamma_{++++} +  \Gamma_{+++-}  + \Gamma_{+-++} +
\Gamma_{+-+-} +  
\Gamma_{++-+}   +  \Gamma_{++--} +  \Gamma_{+--+} +
\Gamma_{+---} . \label{b3} 
\end{equation}
Furthermore,  by   relating   this  sum   to the   forward  scattering
amplitudes, we indeed have a simple way of calculating the temperature
dependent part of the retarded four point amplitude. 

From these low order examples, it is now clear now how to extend these
results to the retarded $n$-point amplitude. 
\begin{equation}
\Gamma_{R}^{(n)}        =       \sum_{i_{k}=\pm}\;\Gamma_{+i_{1}\cdots
i_{n-1}} . \label{b4} 
\end{equation}
Namely, we should keep the index associated with the
largest time vertex fixed to be the physical index ($+$)
and sum over all possible permutations  of the thermal indices.  
It is known\cite{chou:1985es} that this form of the retarded
amplitude is equivalent to the standard one given in terms of multiple
nested commutators.
There
is an even number  of diagrams at every order  and from the properties
already discussed,  the sum  of all diagrams   with more than  one cut
propagator  can   easily  be   seen      to vanish   (graphs    cancel
pairwise). Furthermore, the sum of the single cut propagators can then
be  given a forward   scattering representation which identifies  them
with the imaginary time result as well as makes the evaluation simple. 
By a straightforward generalization of the previous results, we
see that the rule for calculating the general retarded $n$-point
forward scattering amplitudes is as follows 
(compare with Fig. (4)). Let $p_n$ be the
external momenta associated with the latest time vertex. Then, all
propagators whose momenta are flowing towards this vertex are
advanced, whereas the propagators with momenta flowing outward from
the latest time vertex are retarded.

\section{Application}\label{sect4}

As an application of this method,  let us next calculate the $n$-point
photon amplitudes at one loop for  the $1+1$ dimensional massless QED.
It is well  known that in the Schwinger model\cite{schwinger:1962tp}  
only the  self-energy is non vanishing.  
Thus, we are interested mainly in the thermal corrections
to these  amplitudes.   The  simplest,  of   course, is   the retarded
self-energy which can be obtained from the forward scattering 
amplitude diagrams shown in Fig. (1).

\begin{eqnarray}
\Gamma_{R}^{\mu\nu}(p) & = &  - \frac{e^{2}}{2\pi} 
\int {d^{2}k }\;
n_{F}(|k^{0}|)\delta           (k^{2})\nonumber\\           &        &
\hspace{.3in}\times\left({N^{\mu\nu}(k,p)\over
((k+p)^{2}+i\epsilon(k^{0}+p^{0}))} +  {N^{\nu\mu}(k-p,p)\over
((k-p)^{2}-i\epsilon(k^{0}-p^{0}))}\right) . \label{c1} 
\end{eqnarray}
Here, we have 
\begin{eqnarray}
N^{\mu\nu}(k,p)             &            =         &              {\rm
Tr}(\gamma^{\mu}(k\!\!\!\slash+p\!\!\!\slash)
\gamma^{\nu}k\!\!\!\slash
)\nonumber\\   &  =  &  k_{+}^{\mu}  (k+p)_{+}^{\nu}   +
k_{-}^{\mu} (k+p)_{-}^{\nu} 
\end{eqnarray}
and we have defined 
\begin{equation}
k_{\pm}^{\mu} = (\eta^{\mu\nu} \pm \epsilon^{\mu\nu})k_{\nu}\label{c2},
\end{equation}
where $\epsilon^{\mu\nu}$ is the antisymmetric tensor with
$\epsilon^{01}=1$.
It is   clear that, under a  suitable  redefinition of  variables, the
second term in Eq.  (\ref{c1}) becomes identical to  the first and the
integral can be trivially evaluated to give 
\begin{eqnarray}
\Gamma_{R}^{00}(p) & = &     \Gamma_{R}^{11}(p)\nonumber\\ &  =    &
  -\frac{e^{2}}{2\pi}
  \left({1\over    (p^{0}+i\epsilon)-p^{1}}       -      {1\over
  (p^{0}+i\epsilon)+p^{1}}\right)\int                     dk^{1}
  \;{\rm sgn}(k^{1}) n_{F}(|k^{1}|)\nonumber\\   &     =   &
  0\nonumber\\           
\Gamma_{R}^{01}(p)          &       =      &
  \Gamma_{R}^{10}(p)\nonumber\\      &   =   &    
-\frac{e^{2}}{2\pi}\left({1\over
  (p^{0}+i\epsilon)-p^{1}}+{1\over (p^{0}+i\epsilon)+p^{1}}\right)\int
  {dk^{1}}\;{\rm sgn}(k^{1})n_{F}(|k^{1}|)\nonumber\\ &
  = & 0 .
\end{eqnarray}
Similarly, the diagrams for the temperature dependent part of the
retarded three point function are given in Fig (3).  These can also be
evaluated in a simple manner.  Without going into details, we give the
result here, namely, there are only two tensor structures with an
even/odd number of space-like indices. For example,
\begin{eqnarray}
\Gamma_{R}^{000}(p_{1},p_{2},p_{3})   & = & -\frac{e^2}{2\pi}
\left[  \frac{1}{p_{1}^{-}(p_{1}^{-}+p_{2}^{-})} -
        \frac{1}{p_{1}^{+}(p_{1}^{+}+p_{2}^{+})}
+P(p_1,p_2,p_3)\right]
\nonumber\\         &          &       \times\;\int{dk^{1}}\;
{\rm sgn}(k^{1}) n_{F}(|k^{1}|)\nonumber\;\;\; =\;\;\;  0\nonumber\\ 
\Gamma_{R}^{001}(p_{1},p_{2},p_{3}) & = &  -\frac{e^2}{2\pi}
\left[  \frac{1}{p_{1}^{-}(p_{1}^{-}+p_{2}^{-})} +
        \frac{1}{p_{1}^{+}(p_{1}^{+}+p_{2}^{+})}
+P(p_1,p_2,p_3)\right]
\nonumber\\         &           &      \times\;\int{dk^{1}}\;
{\rm sgn}(k^{1}) n_{F}(|k^{1}|)\;\;\; =\;\;\;  0 ,
\end{eqnarray}
where $P(p_1,p_2,p_3)$ represents contributions obtained by cyclic
permutations of the external momenta, and 
$p_i^{\pm}\equiv p_i^0 \pm p_i^1$.
It is to be understood that $p_1^0$ and $p_2^0$ are accompanied by a term
$-i\epsilon$, whereas $p_3^0$ is accompanied by a term $2i\epsilon$.

The calculation of the temperature dependent part of the retarded four
point  amplitude    is straightforward as   well.   
It  corresponds to evaluating the set of diagrams indicated in 
Fig.(4), where the external lines denote, in this case, photon fields. 
Once  again, there  are  only two independent   tensor structures that
arise corresponding  to   whether  there  is an  even/odd   number  of
space-like indices with the values 
\begin{eqnarray}
\Gamma_{R}^{0000}(p_{1},p_{2},p_{3},p_{4})  & = &       -\frac{e^2}{2\pi}
\left[
  \frac{1}{p_{1}^{-}(p_{1}^{-}+p_{2}^{-})(p_{1}^{-}+p_{2}^{-}+p_{3}^{-})} -
\right. \nonumber\\
& {} & \left.\;\;\;\;\;\;\;\;
        \frac{1}{p_{1}^{+}(p_{1}^{+}+p_{2}^{+})(p_{1}^{+}+p_{2}^{+}+p_{3}^{+})}
+P(p_1,p_2,p_3,p_4)\right]
\nonumber\\         &          &       \times\;\int{dk^{1}}\;
{\rm sgn}(k^{1}) n_{F}(|k^{1}|)\nonumber\;\;\;=\;\;\; 0\nonumber\\ 
\Gamma_{R}^{0001}(p_{1},p_{2},p_{3},p_{4}) & = &  -\frac{e^2}{2\pi}
\left[
  \frac{1}{p_{1}^{-}(p_{1}^{-}+p_{2}^{-})(p_{1}^{-}+p_{2}^{-}+p_{3}^{-})} + 
\right. \nonumber\\
& {} & \left.\;\;\;\;\;\;\;\;
 \frac{1}{p_{1}^{+}(p_{1}^{+}+p_{2}^{+})(p_{1}^{+}+p_{2}^{+}+p_{3}^{+})}
+P(p_1,p_2,p_3,p_4)\right]
\nonumber\\         &           &      \times\;\int{dk^{1}}\;
{\rm sgn}(k^{1}) n_{F}(|k^{1}|)\;\;\; =\;\;\;  0 ,
\end{eqnarray}
where the appropriate $i\epsilon$ terms are to be understood.

The pattern  is clear now. As  we calculate  the temperature dependent
parts  of the  retarded  higher  point amplitudes,  only  the external
factors   change  in a   predictable manner   whereas the coefficient,
namely,  the   integral,      does not  change      and  vanishes   by
antisymmetry. Thus, in this model, the  temperature dependent parts of
all the  retarded amplitudes,  at one  loop,  vanish.  This is  indeed
consistent with  the conclusions  of\cite{das:1998gc}  
where  it was shown  that the
temperature dependent part of the effective action in this theory does
not lead to any physical effects.

\section{Further Properties of $1+1$ thermal QED}\label{sect5}

From the results of the previous section, we note  that all the finite
temperature dependent parts of  the  retarded $n$-point amplitudes  in
the $1+1$  dimensional QED vanish to  one  loop order.  This, however,
does not say  anything, in general, about  the higher loop amplitudes.
This is easily seen from  the fact that if we  were to connect any two
external photon lines, the  form of the  zero temperature part of  the
propagator  depends on the type  of  vertices that  it connects.  And,
consequently,  the  pairwise  cancellation  that  took  place  between
diagrams in the   previous section no  longer  holds.  Of course,  the
temperature dependent  part of the  photon propagator,  like any other
propagator, does not  depend on the  type of the vertices it  connects
(namely, for any pair of  thermal indices) and, consequently, leads to
a vanishing  contribution. Thus,  the  problematic diagrams at  higher
loop would  appear to be the  single  cut diagrams  where the internal
photon lines are not cut as in the example shown in 
\hbox{Fig. (5)}.
\vbox{
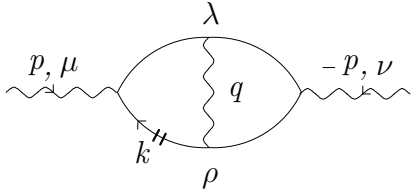
\begin{figure}[h!]
\center{
\setlength{\unitlength}{0.00045833in}
\begingroup\makeatletter\ifx\SetFigFont\undefined%
\gdef\SetFigFont#1#2#3#4#5{%
  \reset@font\fontsize{#1}{#2pt}%
  \fontfamily{#3}\fontseries{#4}\fontshape{#5}%
  \selectfont}%
\fi\endgroup%
{\renewcommand{\dashlinestretch}{30}
\begin{picture}(4662,1971)(0,-10)
\thicklines
\path(1707,474)(1771,598)
\path(1788,433)(1849,554)
\thinlines
\path(562,1117)(562,1027)(472,1027)
\path(202,1027)(203,1026)(207,1023)
	(215,1015)(228,1004)(241,993)
	(254,983)(265,976)(275,971)
	(284,968)(292,967)(300,968)
	(309,971)(319,976)(330,983)
	(343,993)(356,1004)(369,1015)
	(377,1023)(381,1026)(382,1027)
\path(382,1027)(383,1028)(387,1031)
	(395,1039)(408,1050)(421,1061)
	(434,1071)(445,1078)(455,1083)
	(464,1086)(472,1087)(480,1086)
	(489,1083)(499,1078)(510,1071)
	(523,1061)(536,1050)(549,1039)
	(557,1031)(561,1028)(562,1027)
\path(562,1027)(563,1026)(567,1023)
	(575,1015)(588,1004)(601,993)
	(614,983)(625,976)(635,971)
	(644,968)(652,967)(660,968)
	(669,971)(679,976)(690,983)
	(703,993)(716,1004)(729,1015)
	(737,1023)(741,1026)(742,1027)
\path(742,1027)(743,1028)(747,1031)
	(755,1039)(768,1050)(781,1061)
	(794,1071)(805,1078)(815,1083)
	(824,1086)(832,1087)(840,1086)
	(849,1083)(859,1078)(870,1071)
	(883,1061)(896,1050)(909,1039)
	(917,1031)(921,1028)(922,1027)
\path(922,1027)(923,1026)(927,1023)
	(935,1015)(948,1004)(961,993)
	(974,983)(985,976)(995,971)
	(1004,968)(1012,967)(1020,968)
	(1029,971)(1039,976)(1050,983)
	(1063,993)(1076,1004)(1089,1015)
	(1097,1023)(1101,1026)(1102,1027)
\path(1102,1027)(1103,1028)(1107,1031)
	(1115,1039)(1128,1050)(1141,1061)
	(1154,1071)(1165,1078)(1175,1083)
	(1184,1086)(1192,1087)(1200,1086)
	(1209,1083)(1219,1078)(1230,1071)
	(1243,1061)(1256,1050)(1269,1039)
	(1277,1031)(1281,1028)(1282,1027)
\path(22,1027)(23,1028)(27,1031)
	(35,1039)(48,1050)(61,1061)
	(74,1071)(85,1078)(95,1083)
	(104,1086)(112,1087)(120,1086)
	(129,1083)(139,1078)(150,1071)
	(163,1061)(176,1050)(189,1039)
	(197,1031)(201,1028)(202,1027)
\path(2339,579)(2340,580)(2343,584)
	(2351,592)(2362,605)(2373,618)
	(2383,631)(2390,642)(2395,652)
	(2398,661)(2399,669)(2398,677)
	(2395,686)(2390,696)(2383,707)
	(2373,720)(2362,733)(2351,746)
	(2343,754)(2340,758)(2339,759)
\path(2339,759)(2338,760)(2335,764)
	(2327,772)(2316,785)(2305,798)
	(2295,811)(2288,822)(2283,832)
	(2280,841)(2279,849)(2280,857)
	(2283,866)(2288,876)(2295,887)
	(2305,900)(2316,913)(2327,926)
	(2335,934)(2338,938)(2339,939)
\path(2339,939)(2340,940)(2343,944)
	(2351,952)(2362,965)(2373,978)
	(2383,991)(2390,1002)(2395,1012)
	(2398,1021)(2399,1029)(2398,1037)
	(2395,1046)(2390,1056)(2383,1067)
	(2373,1080)(2362,1093)(2351,1106)
	(2343,1114)(2340,1118)(2339,1119)
\path(2339,1119)(2338,1120)(2335,1124)
	(2327,1132)(2316,1145)(2305,1158)
	(2295,1171)(2288,1182)(2283,1192)
	(2280,1201)(2279,1209)(2280,1217)
	(2283,1226)(2288,1236)(2295,1247)
	(2305,1260)(2316,1273)(2327,1286)
	(2335,1294)(2338,1298)(2339,1299)
\path(2339,1299)(2340,1300)(2343,1304)
	(2351,1312)(2362,1325)(2373,1338)
	(2383,1351)(2390,1362)(2395,1372)
	(2398,1381)(2399,1389)(2398,1397)
	(2395,1406)(2390,1416)(2383,1427)
	(2373,1440)(2362,1453)(2351,1466)
	(2343,1474)(2340,1478)(2339,1479)
\path(2339,1479)(2338,1480)(2335,1484)
	(2327,1492)(2316,1505)(2305,1518)
	(2295,1531)(2288,1542)(2283,1552)
	(2280,1561)(2279,1569)(2280,1577)
	(2283,1586)(2288,1596)(2295,1607)
	(2305,1620)(2316,1633)(2327,1646)
	(2335,1654)(2338,1658)(2339,1659)
\path(2339,399)(2338,400)(2335,404)
	(2327,412)(2316,425)(2305,438)
	(2295,451)(2288,462)(2283,472)
	(2280,481)(2279,489)(2280,497)
	(2283,506)(2288,516)(2295,527)
	(2305,540)(2316,553)(2327,566)
	(2335,574)(2338,578)(2339,579)
\put(2293.878,550.123){\arc{2220.110}{3.5846}{4.7585}}
\put(2291.029,1497.134){\arc{2210.905}{1.5220}{2.7013}}
\put(2397.000,550.000){\arc{2220.437}{4.6655}{5.8401}}
\put(2399.843,1497.262){\arc{2211.246}{0.4403}{1.6204}}
\path(4100,1101)(4100,1011)(4190,1011)
\path(3647,1310)(3782,1310)
\path(1622,702)(1529,698)(1528,608)
\path(4460,1011)(4459,1010)(4455,1007)
	(4447,999)(4434,988)(4421,977)
	(4408,967)(4397,960)(4387,955)
	(4378,952)(4370,951)(4362,952)
	(4353,955)(4343,960)(4332,967)
	(4319,977)(4306,988)(4293,999)
	(4285,1007)(4281,1010)(4280,1011)
\path(4280,1011)(4279,1012)(4275,1015)
	(4267,1023)(4254,1034)(4241,1045)
	(4228,1055)(4217,1062)(4207,1067)
	(4198,1070)(4190,1071)(4182,1070)
	(4173,1067)(4163,1062)(4152,1055)
	(4139,1045)(4126,1034)(4113,1023)
	(4105,1015)(4101,1012)(4100,1011)
\path(4100,1011)(4099,1010)(4095,1007)
	(4087,999)(4074,988)(4061,977)
	(4048,967)(4037,960)(4027,955)
	(4018,952)(4010,951)(4002,952)
	(3993,955)(3983,960)(3972,967)
	(3959,977)(3946,988)(3933,999)
	(3925,1007)(3921,1010)(3920,1011)
\path(3920,1011)(3919,1012)(3915,1015)
	(3907,1023)(3894,1034)(3881,1045)
	(3868,1055)(3857,1062)(3847,1067)
	(3838,1070)(3830,1071)(3822,1070)
	(3813,1067)(3803,1062)(3792,1055)
	(3779,1045)(3766,1034)(3753,1023)
	(3745,1015)(3741,1012)(3740,1011)
\path(3740,1011)(3739,1010)(3735,1007)
	(3727,999)(3714,988)(3701,977)
	(3688,967)(3677,960)(3667,955)
	(3658,952)(3650,951)(3642,952)
	(3633,955)(3623,960)(3612,967)
	(3599,977)(3586,988)(3573,999)
	(3565,1007)(3561,1010)(3560,1011)
\path(3560,1011)(3559,1012)(3555,1015)
	(3547,1023)(3534,1034)(3521,1045)
	(3508,1055)(3497,1062)(3487,1067)
	(3478,1070)(3470,1071)(3462,1070)
	(3453,1067)(3443,1062)(3432,1055)
	(3419,1045)(3406,1034)(3393,1023)
	(3385,1015)(3381,1012)(3380,1011)
\path(4640,1011)(4639,1012)(4635,1015)
	(4627,1023)(4614,1034)(4601,1045)
	(4588,1055)(4577,1062)(4567,1067)
	(4558,1070)(4550,1071)(4542,1070)
	(4533,1067)(4523,1062)(4512,1055)
	(4499,1045)(4486,1034)(4473,1023)
	(4465,1015)(4461,1012)(4460,1011)
\put(270,1296){\makebox(0,0)[lb]{\smash{{{\SetFigFont{12}{14.4}{\rmdefault}{\mddefault}{\itdefault}p}}}}}
\put(443,1229){\makebox(0,0)[lb]{\smash{{{\SetFigFont{12}{14.4}{\rmdefault}{\mddefault}{\itdefault},}}}}}
\put(637,1266){\makebox(0,0)[lb]{\smash{{{\SetFigFont{12}{14.4}{\rmdefault}{\mddefault}{\updefault}$\mu$}}}}}
\put(4058,1234){\makebox(0,0)[lb]{\smash{{{\SetFigFont{12}{14.4}{\rmdefault}{\mddefault}{\itdefault},}}}}}
\put(4253,1233){\makebox(0,0)[lb]{\smash{{{\SetFigFont{12}{14.4}{\rmdefault}{\mddefault}{\updefault}$\nu$}}}}}
\put(3862,1289){\makebox(0,0)[lb]{\smash{{{\SetFigFont{12}{14.4}{\rmdefault}{\mddefault}{\itdefault}p}}}}}
\put(2551,986){\makebox(0,0)[lb]{\smash{{{\SetFigFont{12}{14.4}{\rmdefault}{\mddefault}{\itdefault}q}}}}}
\put(2265,0){\makebox(0,0)[lb]{\smash{{{\SetFigFont{12}{14.4}{\rmdefault}{\mddefault}{\updefault}$\rho$}}}}}
\put(2257,1806){\makebox(0,0)[lb]{\smash{{{\SetFigFont{12}{14.4}{\rmdefault}{\mddefault}{\updefault}$\lambda$}}}}}
\put(1484,219){\makebox(0,0)[lb]{\smash{{{\SetFigFont{12}{14.4}{\rmdefault}{\mddefault}{\itdefault}k}}}}}
\end{picture}
}
}
\bigskip
   \label{fig4}
\caption{A typical two-loop self-energy graph with a single fermion line cut.}
  \end{figure}
}
In this simple theory, however,  things are rather special. Let  us
illustrate this with a general two loop diagram which would correspond
to connecting  a  pair  of external photon   lines  in the  four point
vertex functions discussed in  Sec. \ref{sect3}.  
First,  we note that  in the  $1+1$
dimensional QED,   the photon becomes massive  and   as a  result, the
photon propagator in the Feynman gauge would have the form 
\begin{equation}
D_{\lambda\rho}(q)  =        {\eta_{\lambda\rho}\over    (q^{2}   
-     m^{2} +
i\epsilon)}, \label{d1} 
\end{equation}
where  the  photon mass  can    be identified with  $m^{2}={e^{2}\over
\pi}$. 

The general form for a diagram 
with all \lq\lq +'' vertices in the $2n$-point amplitude,
at one loop, is easily seen to be (in this theory, only the even point
amplitudes are nonzero by charge conjugation invariance) 
\begin{eqnarray}
T^{\mu_{1}\cdots    \mu_{2n}}    &    \sim    &   \int    {d^{2}k\over
(2\pi)^{2}}\left[k_{+}^{\mu_{1}}(k+p_{1})_{+}^{\mu_{2}}\cdots
(k+\cdots    +p_{2n-1})_{+}^{\mu_{2n}}\right.\nonumber\\    &        &
\hspace{.5in}  \left.  +  k_{-}^{\mu_{1}}(k+p_{1})_{-}^{\mu_{2}}\cdots
(k+\cdots   +p_{2n-1})_{-}^{\mu_{2n}}\right]\nonumber\\  &   &  \times
{n_{F}(|k^{0}|)\delta(k^{2})\over    ((k+p_{1})^{2}+i\epsilon)  \cdots
((k+\cdots p_{2n-1})^{2}+i\epsilon)} . 
\end{eqnarray}
If  we now   connect  any photon  line  (and  consider  only the  zero
temperature  part of the photon propagator because as   we    have argued 
before  the    finite
temperature parts would cancel pairwise among the  sum of the diagrams
in  the combination giving the  retarded amplitude), then we trivially
obtain 
\begin{equation}
\int \frac{d^2 q}{(2\pi)^2}
T^{\mu_{1}\cdots \mu_{i} \cdots \mu_{j} \cdots \mu_{2n}}\;
{\eta_{\mu_{i}\mu_{j}}\over (q^{2}  -
m^{2} +i\epsilon)} = 0\label{d2} .
\end{equation}
Here  we have assumed that  the photon line  being connected carries a
momentum  $q$ and  the vanishing of  the amplitude  in Eq.  (\ref{d2})
follows  from the fact that for  any two  arbitrary vectors, 
$A^{\lambda}$ and $B^{\rho}$, 
\begin{equation}
A_{\pm}^{\lambda}B_{\pm}^{\rho}\;\eta_{\lambda\rho} = 0 . \label{d3} 
\end{equation}

A similar argument shows that,  in this simple  
model, each diagram with an internal
photon   line identically  vanishes  and,  consequently,  the retarded
amplitude continues to vanish even at higher loops.  However, it is to
be emphasized  that (\ref{d3}) only leads 
to the  vanishing of the diagram, provided the integral  
in (\ref{d2}) is convergent. 
As  is well known, the   finite temperature integrals are
ultraviolet finite  and   hence there is   no  such  problem  from the
ultraviolet region. For a   massive  photon, similarly, there   is  no
infrared  problem either and hence  the diagram, indeed, vanishes.  On
the other hand, if we did not include the photon mass, these integrals
would be infrared divergent and, consequently, we cannot conclude that
each of  these diagrams vanishes  individually. In fact, they  don't and
only when we sum the amplitude to all orders, all such divergent terms
would cancel as (\ref{d2}) shows. 

The discussion, so  far,  would seem  to  say that when two   external
photon lines are contracted,  the  diagrams individually vanish for  a
single cut fermion  line in the  Feynman gauge.  This is, however, not
true in a general gauge. Without going into details, let us simply
summarize  here    that the $q^{\lambda}q^{\rho}$ terms in   the  photon
propagator   do   contribute graph by     graph.   However, when their
contribution is summed for a fixed single cut fermion line, the effect
is to lead to an integral  which vanishes by anti-symmetry (the number
of alternating step functions becomes odd).  This could also have been
inferred from  the gauge  invariance  of this model. Namely,  we  know
that, in  this model,   any  temperature dependent correction to   any
amplitude must be gauge  invariant and, consequently, if  an amplitude
vanishes  in  the  Feynman gauge,  it  must  also vanish in  any other
gauge. This model, in this sense,  is special and immediately leads to
the fact that in this $1+1$ dimensional QED, the temperature dependent
corrections to all  the retarded amplitudes  vanish to all loops. This
is, indeed, consistent with the conclusions of\cite{das:1998gc}.

\section{Conclusion}\label{sect6}

In this paper, we have  identified, in a simple diagrammatic way,
the unique  combination of $n$-point   causal amplitudes in the  real time
formalism  that corresponds   to the retarded $n$-point amplitude
obtained by analytic continuation
from the imaginary time formalism. We have also
given a  simple method of  calculating the temperature dependent parts of
the retarded   $n$-point amplitudes, at   least to one loop  order, by
identifying them with   the forward  scattering amplitude
of on-shell thermal particles. 
(The extension of this approach to higher orders will be
reported elsewhere.)
As an  application, we have  calculated and
shown that  all the  temperature   dependent  parts of  the   retarded
$n$-point amplitudes for $1+1$ dimensional  massless QED vanish to one
loop.  For   this simple model,  it  turns  out  that  the temperature
dependent  parts of all  the retarded $n$-point amplitudes also vanish
to all loops. 

\acknowledgements   

A.D.   would like to thank   the  members of the Mathematical  Physics
Department of USP  for hospitality where  this work was  done. A.D. is
supported  in  part    by US  DOE    Grant number  DE-FG-02-91ER40685,
NSF-INT-9602559 and FAPESP.  F.T.B, J.F.  and  A.J.S. are partially
supported by CNPq (the National Research Council of Brazil) and 
F.T.B is supported in part by ``Programa de Apoio a N\'ucleos de
Excel\^encia'' (PRONEX).


\begin{thebibliography}{99}

\bibitem{gross:1981br}
D.~J. Gross, R.~D. Pisarski, and L.~G. Yaffe, Rev. Mod. Phys. {\bf 53},  43
  (1981).

\bibitem{weldon:1982aqweldon:1983jn}
H.~A. Weldon, Phys. Rev. {\bf D26},  1394  (1982);
{\bf D28},  2007  (1983).

\bibitem{kajantie:1985xx}
K. Kajantie and J. Kapusta, Ann. Phys. {\bf 160},  477  (1985).

\bibitem{braaten:1990mzbraaten:1990az}
E. Braaten and R.~D. Pisarski, Nucl. Phys. {\bf B337},  569  (1990)
; {\bf B339},  310  (1990);\\ Phys. Rev.
{\bf D45}, 1827 (1992).

\bibitem{smilga:1996cm}
A. V. Smilga, {\em Physics of thermal QCD} Phys. Rept.,
{\bf 291}, 1-106 (1997).

\bibitem{kapusta:book89}
J.~I. Kapusta, {\em Finite Temperature Field Theory} (Cambridge University
  Press, Cambridge, England, 1989).

\bibitem{lebellac:book96}
M.~L. Bellac, {\em Thermal Field Theory} (Cambridge University Press,
  Cambridge, England, 1996).

\bibitem{das:book97}
A. Das, {\em Finite Temperature Field Theory} (World Scientific, NY, 1997).

\bibitem{frenkel:1991tsbrandt:1997se}
J. Frenkel and J.~C. Taylor, Nucl. Phys. {\bf B374},  156  (1992).
\\
F.~T. Brandt and J. Frenkel, Phys. Rev. {\bf D56},  2453  (1997).


\bibitem{landsman:1987uw}
N.~P. Landsman and C.~G. van Weert, Phys. Rept. {\bf 145},  141  (1987).


\bibitem{evans:1992kyevans:1993ak}
T.~S. Evans, Nucl. Phys. {\bf B374},  340  (1992);
Phys. Rev. {\bf D47},  4196  (1993).


\bibitem{baier:1994yh}
R. Baier and A. Niegawa, Phys. Rev. {\bf D49},  4107  (1994).

\bibitem{carrington:1996rx}
M.~E. Carrington and U. Heinz, Eur. Phys. J. {\bf C1},  619  (1998).

\bibitem{chou:1985es}
Kuang-chao Chou, Zhao-bin Su, Bai-lin Hao and Lu Yu,
Phys. Rept. {\bf 118}, 1 (1985).

\bibitem{schwinger:1962tp}
J. Schwinger, Phys. Rev. {\bf 128},  2425  (1962).

\bibitem{das:1998gc}
A. Das and A.~J. da~Silva, hep-th/9808027  (1998).

\end{thebibliography}
\end{document}